\documentclass[aps,floats,amssymb,amsmath,prb,nofootinbib,twocolumn,superscriptaddress]{revtex4-1}

\usepackage{booktabs}
\usepackage{calc}
\usepackage{psfrag}
\usepackage{graphicx}
\usepackage{color}
\usepackage[dvipsnames]{xcolor}
\usepackage[unicode=true,pdfusetitle,
 bookmarks=true,bookmarksnumbered=false,bookmarksopen=false,
 breaklinks=false,pdfborder={0 0 0},backref=false,colorlinks=true]
 {hyperref}
\usepackage{geometry}
\pdfpageattr {/Group << /S /Transparency /I true /CS /DeviceRGB>>} 

\usepackage{pdfpages}

\makeatletter
\AtBeginDocument{\let\LS@rot\@undefined}
\makeatother

\begin{document}

\title{Real-frequency Diagrammatic Monte Carlo at Finite Temperature}
\author{J. Vu\v ci\v cevi\'c}
\affiliation{Scientific Computing Laboratory, Center for the Study of Complex Systems, Institute of Physics Belgrade,
University of Belgrade, Pregrevica 118, 11080 Belgrade, Serbia}
\author{M. Ferrero}
\affiliation{CPHT, CNRS, Ecole Polytechnique, Institut Polytechnique de Paris, Route de Saclay, 91128 Palaiseau, France}
\affiliation{Coll\`ege de France, 11 place Marcelin Berthelot, 75005 Paris, France}

\begin{abstract}
Diagrammatic expansions are a central tool for treating correlated electron
systems.  At thermal equilibrium, they are most naturally defined within the
Matsubara formalism.
However, extracting any dynamic response function from a Matsubara calculation
ultimately requires the ill-defined analytical continuation from the imaginary-
to the real-frequency domain.  It was recently proposed [Phys. Rev. B 99,
035120 (2019)] that the internal Matsubara summations of any
interaction-expansion diagram can be performed analytically by using symbolic
algebra algorithms.  The result of the summations is then an analytical
function of the complex frequency rather than Matsubara frequency.  Here we
apply this principle and develop a diagrammatic Monte Carlo technique which
yields results directly on the real-frequency axis.  We present results for the
self-energy $\Sigma(\omega)$ of the doped 32x32 cyclic square-lattice Hubbard
model in a non-trivial parameter regime, where signatures of the pseudogap
appear close to the antinode.
We discuss the behavior of the perturbation series on the real-frequency axis and in
particular show that one must be very careful when using the maximum entropy
method on truncated perturbation series.
Our approach holds great promise for future application in cases when
analytical continuation is difficult and moderate-order perturbation theory
may be sufficient to converge the result.
\end{abstract}

\maketitle

\section{Introduction}

Interacting lattice-fermions are one of the central subjects in condensed
matter theory.  Especially in two dimensions, a full many-body solution for
even the simplest models (e.g. the Hubbard model), is a formidable task.  In
recent decades, great progress has been achieved using Monte Carlo algorithms
for the summation of various diagrammatic
expansions.
The main advantage of this
approach is that the approximations can be controlled, i.e. convergence of the
results with respect to the control parameters can be systematically verified.
The control parameters of the calculations are most commonly the lattice size and
the maximal perturbation order. Some algorithms~\cite{rubtsov2004,rubtsov_prb_2005,werner_prl_2006,werner_prb_2006,gull_epl_2008,gullRMP2011,profumoPRB2015,moutenet2019,bertrand2019a,bertrand2019b,cohen2014a,cohen2014b,cohenPRL2015,boag2018,edelstein2019,vucicevicPRL2019} are very efficient
for small systems but have not yet reached very large lattice sizes, while
others~\cite{Prokofev1998,Prokofev2007,vanhoucke_2010,Bourovski2004,riccardo_2017,moutenet2018,fedor_2017,wuPRB2017}
can address the thermodynamic limit directly but are limited in the
number of perturbation orders that can be computed.

In thermal equilibrium, expansions are naturally formulated within the Matsubara formalism, with all the propagators defined in imaginary time/frequency.
Therefore, to obtain dynamic response functions, one needs to perform the analytical continuation from the imaginary to the real frequency domain.
This procedure is notoriously ill-defined and becomes especially difficult when the Matsubara axis data contain statistical noise, as is the case with all Monte Carlo results.
The problem is further exacerbated with increasing temperature.
As the discrete imaginary Matsubara frequencies spread out and move away from the real axis, the statistical noise chips away more and more information from the Matsubara data.
The most common way of analytically continuing a noisy result is the maximum entropy method (MEM)\cite{macridin2004,levycpc2017}, but it requires ``the default model'', an a priori qualitative knowledge of the real-frequency spectrum that may not always be available; it is difficult to control and estimate the error bars of any such procedure.

Analytical continuation is a common hurdle in finite-temperature calculations, and it came up recently in the study of transport in the optical lattice realizations of the Hubbard model\cite{huang2018,vucicevicPRL2019}. 
It turns out that the direct-current resistivity is particularly difficult to extract from the imaginary-axis current-current correlation function.
But even the self-energy is often interpreted only on the imaginary axis\cite{wuPRB2017}, as analytical continuation is considered ultimately unreliable.
This particularly hinders the progress in the study of the pseudo gap phase and superconductivity in the cuprates, where one would like to compare the momentum-resolved spectral function to experiments\cite{staarPRB2014,gullPRB2015}.
The ability to reliably calculate the spectral function becomes even more important in the view of the recent photoemission measurements (ARPES) in the cold atom realizations of the Hubbard model\cite{brown2019}.

There are alternative routes that avoid analytical continuation altogether
(Keldysh formalism~\cite{aokiRMP2014,profumoPRB2015,moutenet2019,bertrand2019a,bertrand2019b,cohen2014a,cohen2014b,cohenPRL2015,boag2018}, exact diagonalization techniques~\cite{Rigol2006,kokaljPRB2017,brownScience2018,vucicevicPRL2019}), but those have so far been limited to impurity models or small lattice sizes.
It is therefore of primary importance to try and develop methods that avoid the analytical continuation, but are not limited by lattice size.

As was recently proposed~\cite{taheridehkordiPRB2019}, an opportunity lies in symbolic algebra algorithms. 
One can implement a recursive transformation to perform analytically all the internal Matsubara frequency summations for any interaction-expansion diagram, for any quantity.
The result of the Matsubara frequency summations is an analytical expression for the contribution of a given diagram to the given dynamic quantity, in the whole of the complex-frequency plane, rather than just in the discrete set of points along the imaginary axis.
The general idea is, however, not entirely new - at perturbation order 2, the Matsubara summations for the self-energy diagrams can be carried out by hand, which leads to the well known real-axis iterative perturbation theory (RAIPT)\cite{kajueterPRL1996,pothoffPRB1997,dasariEPJB2016}.
Similarly, the bubble diagrams can be easily rewritten in terms of real frequencies, which has applications in the GW method\cite{onufrievaPRB2002,onufrievaPRL2009,onufrievaPRL2012} and the calculation of optical conductivity within Kubo formalism\cite{terletskaPRL2011,vucicevicPRB2013,vucicevicPRL2015}.
In the context of diagrammatic Monte Carlo, however, obtaining the analytical expression for each diagram of interest is only a part of the problem. 
In fact, there are several immediate obstacles in applying the algorithmic Matsubara summations in a calculation of quantities at perturbation order $\geq 3$.

Here we address these problems and successfully develop and test a diagrammatic
Monte Carlo technique that yields results directly on the real-frequency axis,
yet can treat very large systems.  We present solutions for the
momentum-resolved self-energy for a doped $32\times 32$ Hubbard lattice, in a
non-trivial parameter regime where results are almost converged at order 5. Our
results show that in this regime precursor signatures of the pseudogap are
visible in the real-frequency antinodal self-energy. We also show that the
truncation of the perturbation series leads to non-causal features that
challenge the use of MEM to obtain real-frequency data from Matsubara axis
results.

\section{Model}
We solve the Hubbard model on the square lattice
\begin{equation}
 H = -t\sum_{\sigma, \langle i,j\rangle} c^\dagger_{\sigma i} c_{\sigma j}+U\sum_{i} n_{\uparrow i}n_{\downarrow i}-\mu\sum_{\sigma, i} n_{\sigma i},
\end{equation}
where $c^\dagger_{\sigma i}/c_{\sigma i}$ create/annihilate an electron of spin $\sigma$ at the lattice site $i$. The hopping amplitude between the nearest neighbors is denoted $t$, and we set $D=4t$ as the unit of energy. 
The density operator is $n_{\sigma i}=c^\dagger_{\sigma i}c_{\sigma i}$, the chemical potential $\mu$, and the on-site Hubbard interaction $U$.
We restrict to paramagnetic solutions with full lattice symmetry.

\section{Method}
\subsection{Symbolic algorithm}
Following similar steps as those in Ref.~\cite{taheridehkordiPRB2019},
we first define the Hartree-shifted bare Green's function of the model
$
  G^\mathrm{HF}_{0,\mathbf{k}}(i\omega)=[i\omega-\varepsilon(\mathbf{k})]^{-1}
$
where we absorbed the chemical potential and the Hartree shift in the dispersion $\varepsilon(\mathbf{k})$, i.e.
\begin{equation}
  \varepsilon(\mathbf{k})=-\mu+Un_\sigma-2t(\cos k_x + \cos k_y) 
\end{equation}
where $\mathbf{k}=(k_x,k_y)$ is the momentum.
For the sake of clarity we omit the integer index $n$ in the fermionic Matsubara frequency, $i\omega\equiv i\omega_n=i(2n+1)\pi T$, where $T$ is temperature.
We reserve the subscript in $i\omega$ for denoting different Matsubara variables.
We denote $n_\sigma$ the density per spin evaluated in the interacting problem.

The self-energy $\Sigma$ can be written as a series in the interaction amplitude $U$
\begin{eqnarray}  \label{eq:total}
 \Sigma_\mathbf{k}(i\omega) 
           &=&  \sum_{N=1}^{\infty} (-U)^N \sum^{{\cal N}_N}_{\alpha=1} {\cal D}^{N,\alpha}_{\mathbf{k}}(i\omega)
\end{eqnarray}
where $N$ is the perturbation order, ${\cal N}_N$ is the number of distinct diagrams in the given expansion, $\alpha$ enumerates the diagrams, and ${\cal D}^{N,\alpha}_{\mathbf{k}}$ is the contribution of $\alpha$'th diagram in the $N$'th order. If the diagrams are written in terms of the Hartree-shifted bare propagator there is no need for tadpole insertions in the topology of the diagrams (see Appendix \ref{sec:diagram_topologies}). 

The contribution of a general diagram to the bare series for self-energy written in terms of $G^\mathrm{HF}_{0,\mathbf{k}}(i\omega_n)$ is given by
\begin{eqnarray} \nonumber
 && {\cal D}^{N,\alpha}_\mathbf{k}(i\omega) = \\ \nonumber
 &&(-1)^{N_b} \sum_{\substack{\mathbf{k}_1..\mathbf{k}_{M} \\ i\Omega_1..i\Omega_M}} \prod_\gamma \frac{1}{
 \sum_{(s,j)\in {\cal K}_\gamma} s \, i\Omega_j - \varepsilon\big({\sum_{(s,j)\in {\cal K}_\gamma} s \, \mathbf{k}_j} \big) } \\ \label{eq:diag_contrib}
\end{eqnarray}
$N_b\equiv N_b^{N,\alpha}$ is the number of fermionic loops (bubbles) in the given diagram: each bubble carries one independent fermionic frequency and momentum.
Each interaction carries a bosonic frequency $i\nu\equiv i\nu_n=2n\pi T$ and momentum, but some are not independent due to conservation laws.
We denote $M$ the total number of independent degrees of freedom, each consisting of a frequency and momentum $(i\Omega_j,\mathbf{k}_j)$, where $i\Omega$ can be either fermionic or bosonic.
There are $2N-1$ Green's functions in each diagram, indexed by $\gamma$.
Each Green's function depends on a certain subset of the internal degrees of freedom and possibly the external variables, indexed $j\in[0,M]$ (we take $\mathbf{k}_0\equiv\mathbf{k}$, $i\Omega_0\equiv i\omega$), and each entering with a sign $s=\pm 1$ in the corresponding sums. 
The Green's function $\gamma$ is fully defined by a set of sign/index pairs ${\cal K}_\gamma\equiv{\cal K}_\gamma^{N,\alpha}$. The Green's functions may not be unique, i.e. it is possible that ${\cal K}_\gamma={\cal K}_{\gamma'}$. For the discussion of the Feynmann rules leading to the general expression Eq.~\ref{eq:diag_contrib}, we refer the reader to the classic textbook Ref.~\onlinecite{richardmattuck1992}. For a worked out example of Eq.~\ref{eq:diag_contrib} in the 4th order of perturbation, see Appendix~\ref{sec:eq_example}.

As a function of any given internal Matsubara frequency $i\Omega_c$, and for a fixed choice of the remaining internal and external degrees of freedom, the contribution to self-energy from any given diagram $(N,\alpha)$ has the form of a product of poles
\begin{equation} \label{eq:product_of_poles}
 {\cal D}_\mathbf{k}(i\omega) = (-1)^{N_b}\sum_{\substack{\mathbf{k}_1..\mathbf{k}_{M} \\ \{i\Omega_j\}_{j\neq c}}} P \sum_{i\Omega_c} \prod_\gamma \frac{1}{(i\Omega_c-z_\gamma)^{m_\gamma}}
\end{equation}
where $P$ and $z_\gamma$ implicitly depend on the rest of the internal and external variables, and here we assume that $\gamma$ goes only over the unique Green's functions that depend on the given $i\Omega_c$, and $m_\gamma \in \mathbb{N}$ is the number of appearances of the $\gamma$'th Green's function in the diagram.
Using the partial fraction expansion, and an analytic expression for the derivative of a product of an arbitrary number of poles (see Appendix \ref{sec:eq6}), we can perform the transformation
\begin{eqnarray} \label{eq:main_transformation}
&&\prod_\gamma \frac{1}{(z-z_\gamma)^{m_\gamma}} = 
  \sum_\gamma \sum_{r=1}^{m_\gamma} \frac{1}{(z-z_\gamma)^{r}} \times \\ \nonumber
&& \;\;\;\;\;\;\;\;\;\; \times(-1)^{m_\gamma-r} \sum_{{\cal C}\{p_{\gamma'\neq\gamma} \in {\mathbb N_0}\}:\sum_{\gamma'\neq\gamma} p_{\gamma'} = m_\gamma-r} \times \\ \nonumber 
&& \;\;\;\;\;\;\;\;\;\; \times \prod_{\gamma'\neq\gamma}\frac{(m_{\gamma'}+p_{\gamma'}-1)!}{p_{\gamma'}!(m_{\gamma'}-1)!} \frac{1}{(z_\gamma-z_\gamma')^{m_{\gamma'}+p_{\gamma'}}}
\end{eqnarray}
Here $\cal C...$ denotes all combinations of a non-negative-integer $p$-per-pole $\gamma'\neq\gamma$, such that the total sum of $p$'s is equal $m_\gamma -r$.
Therefore, after selecting one internal Matsubara variable, the full expression can be rewritten as a sum of poles in that Matsubara variable. Then, one may proceed to perform the Matsubara summation of each term using
\begin{equation}\label{eq:mats_sum}
 \sum_{i\Omega} \frac{1}{(i\Omega - z)^r} = -\frac{\eta}{(r-1)!} \partial^{r-1} n_\eta(z)
\end{equation}
with $\eta=\pm 1$ for bosonic/fermionic Matsubara frequency.
$n_\eta$ is the Bose/Fermi distribution function.
Here we can immediately get rid of the complex part of $z$ because
\begin{equation} \label{eq:remove_imaginary_in_distrib_arg}
  \partial^{r}_\omega n_\eta(\omega+i\Omega_{\eta'}) = \eta'\partial^{r}_\omega n_{\eta'\cdot\eta}(\omega)
\end{equation}
where $\eta'=-1$ or $+1$ denotes whether $i\Omega_{\eta'}$ is fermionic or bosonic Matsubara frequency, respectively.
Note that the derivatives $\partial^r n$ can be expressed analytically for the purpose of precise numerical evaluation (details in Appendix \ref{sec:eq_distribs}).

Now the remaining Matsubara variables appear only in the denominators of fractions which can again be interpreted as poles with respect to these variables, and the procedure can be applied recursively until we have gotten rid of all the Matsubara variables. For a detailed example of the symbolic algorithm and an illustration of Eq.~\ref{eq:product_of_poles}, see Appendix~\ref{sec:eq_example}.

\begin{figure*}[ht]
\includegraphics[width=4.9in, trim=1.95cm 0 2.7cm 0, clip,]{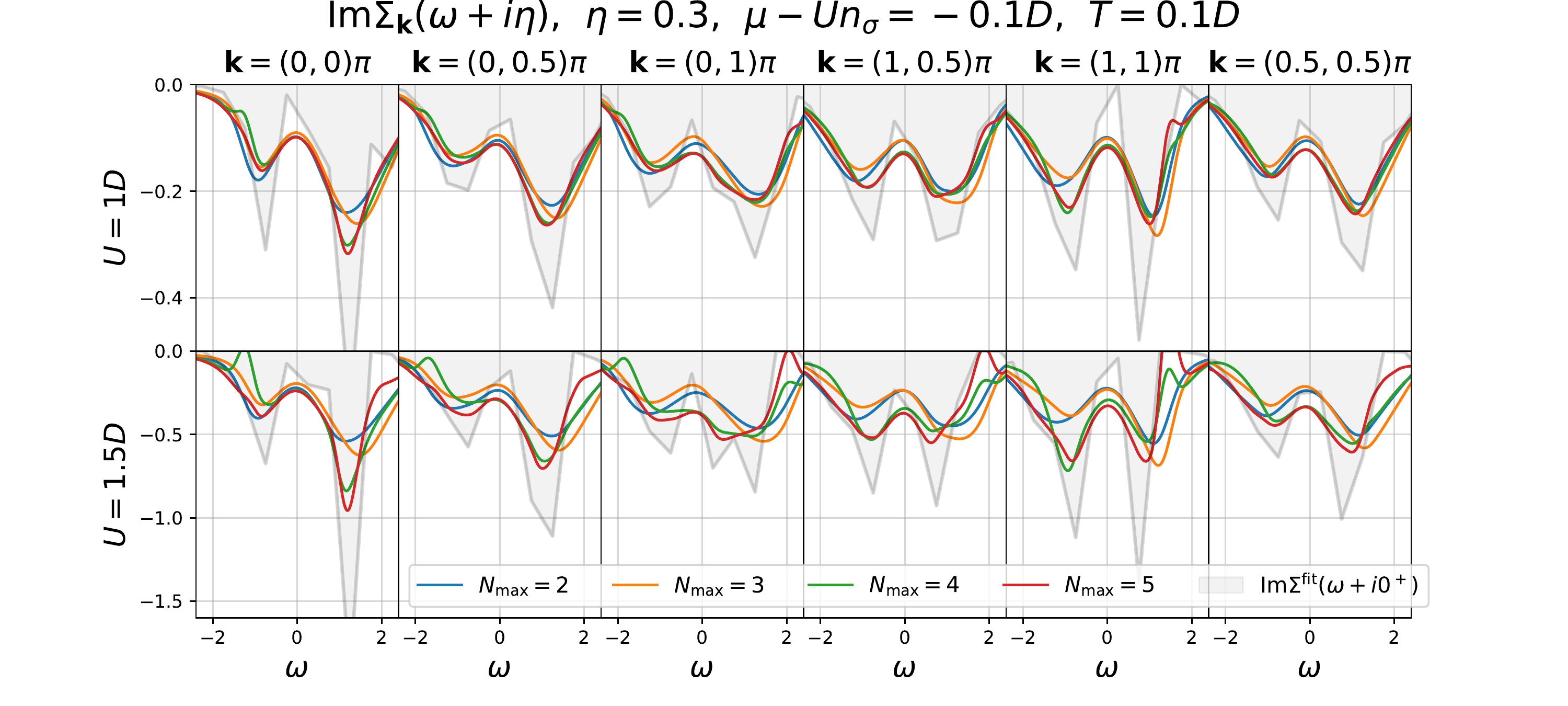}
\includegraphics[width=1.5in, trim=0.2cm 0 0cm 0, clip]{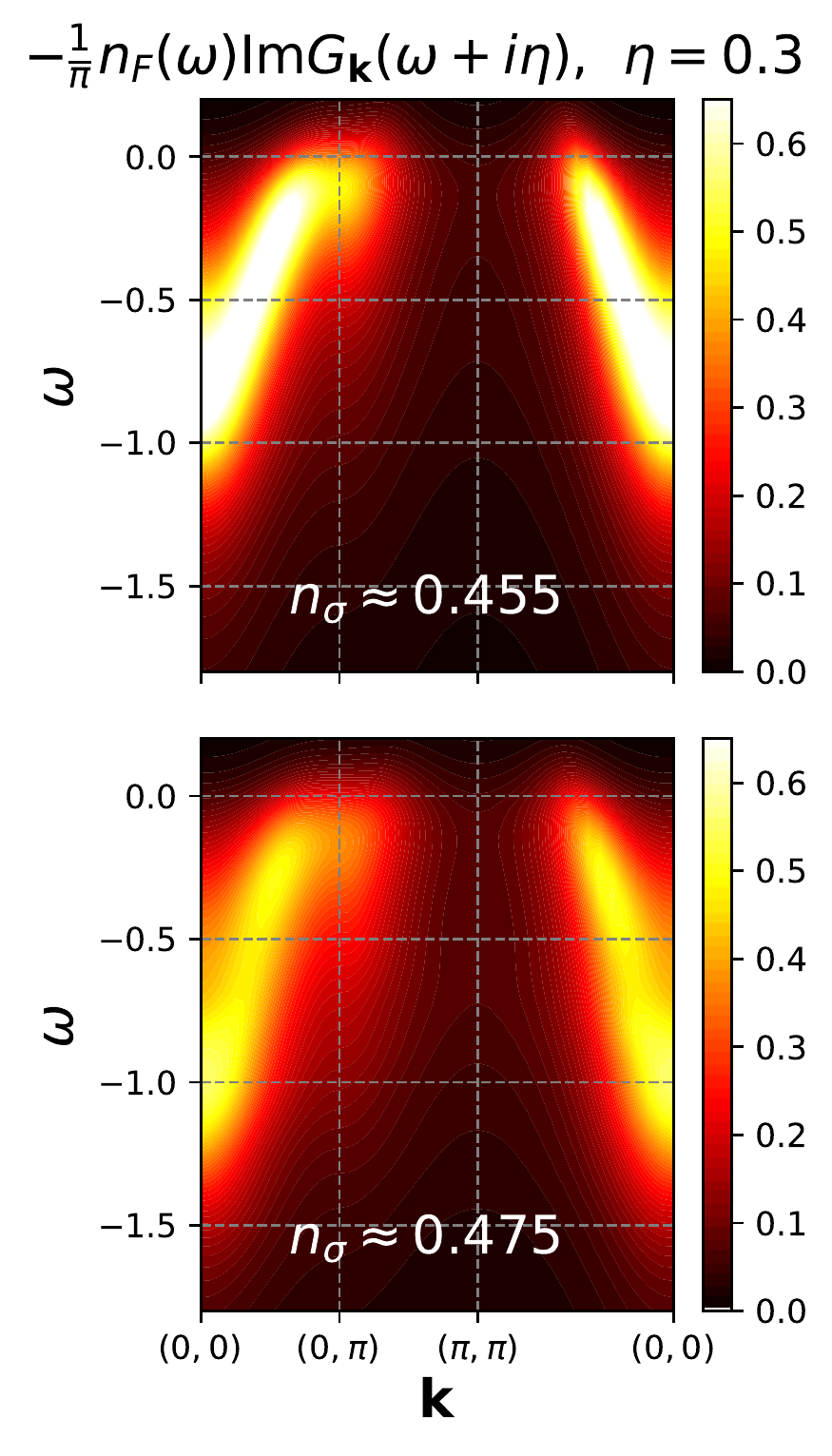}
\caption{
Calculation for the 32x32 Hubbard lattice at two values of $U$, $T=0.1D$,$\mu-Un_\sigma=0.1D$. These parameters correspond to densities per spin indicated in rightmost panels, i.e. dopings $\delta(U=1)\approx 9\%$ and $\delta(U=1.5)\approx 5\%$. Left: Imaginary part of the self-energy $\mathrm{Im}\Sigma(\omega+i\eta)$, at a distance $\eta=0.3D$ from the real axis, for various $\mathbf{k}$ vectors. Different lines correspond to different maximal perturbation orders in the calculation, $N_\mathrm{max}$. Gray-shaded curve is the piecewise-trapezopid fit at $\eta=i0^+$, obtained with resolution $\Delta\omega=1.6\eta$.
Right: the corresponding filled part of the spectral function, broadened with $\eta$, and interpolated in $\mathbf{k}$-space. The result is obtained with $5.12\times10^7$ Monte Carlo steps per diagram.
}
\label{fig:32x32w}
\end{figure*}

The final result has the form of a sum of poles on the real axis
\begin{equation}\label{eq:sum_of_poles}
 {\cal D}_\mathbf{k}(z) = (-1)^{N_\mathrm{b}}\sum_{\mathbf{k}_1...\mathbf{k}_M}\sum_\kappa \frac{A_\kappa}{(z-\omega_\kappa)^{m_\kappa}}
\end{equation}
with $\omega_\kappa = \sum_\gamma s^\kappa_\gamma \varepsilon({\sum_{(s,j)\in{\cal K}^\kappa_\gamma} s\mathbf{k}_j})$, which is a series of terms equal up to the sign $s^\kappa_\gamma=\pm1$ to the dispersion $\varepsilon$, evaluated at various possible linear combinations of the internal/external momenta, as they appear in the Green's functions (indexed $\gamma$). The series can be of any length $\leq2N-1$ and include an arbitrary subset of $\gamma$'s.
The amplitude for each (unique) pole $(\omega_\kappa,m_\kappa)$ is given by a large sum of terms of the general form
\begin{equation}\label{eq:amp_final}
 A_\kappa = \sum_\varsigma \frac{a_{\varsigma}}{b_{\varsigma}}\prod_\zeta \frac{1}{\omega_{\zeta\varsigma}^{m_{\zeta\varsigma}}} \prod_\varrho \partial^{r_{\varrho\varsigma}} n_{\eta_\varrho}(\omega_{\varrho\varsigma})
\end{equation}
$a,b$ are integers, $m$ positive integers. 
$\omega_{\zeta\varsigma}$ and $\omega_{\varrho\varsigma}$ have the same general form as $\omega_\kappa$, but do not necessarily coincide with any of the $\omega_\kappa$'s, and may differ from one another.
The products over $\zeta$ and $\varrho$ may be of various lengths including 0.
$\omega$'s (and thus $A_\kappa$'s) are implicitly dependent on the internal and external momenta.

The symbolic forms for $A_\kappa$ and $\omega_\kappa$ need be obtained only once for any given diagram, independently of the choice of the lattice geometry, parameters of the Hamiltonian, or temperature. See Appendix~\ref{sec:poles_and_terms_stats} for numbers of poles $\omega_\kappa$ and terms in $A_\kappa$ at various perturbation orders.

\subsection{Application in diagrammatic Monte Carlo}
Evaluating the prefactor $A_\kappa$ numerically is not straightforward for several reasons.

First, the terms in $A_\kappa$ containing at least one ratio $1/\omega^m$ or a bosonic $\partial^r n_\eta(\omega)$ will diverge if the corresponding $\omega$ goes to zero.
For any finite lattice this will occur regularly during the Monte Carlo sampling, but even in the thermodynamic limit, $\omega$ can approach arbitrarily close to zero.
Our solution for this problem is to add small shifts to a certain choice of $\varepsilon$'s appearing in $\omega$.
This is done at the symbolic level, in a way that $|\omega|$ can never be smaller than a given value that we set to be $\sim10^{-10}-10^{-6}$ depending on the perturbation order.
Note that even this will cause the terms in $A_\kappa$ to be very large by absolute value (order as large as $10^{30}$), yet they will cancel to produce contributions to $A_\kappa$ of order  $\lesssim 1$. This greatly exceeds the capability of standard precision arithmetic which handles only around $\sim 16$ digits.
We have found the solution in using multiple precision floating point types which can store more digits and allow for subtraction of large numbers, as required in our algorithm. The additional approximation made by numerical shifts can be controlled, and we have checked on several examples that the result is insensitive to the precise choice of the numerical parameters (size of the shifts and the choice of the floating point precision). Surely, the shifts can be always made smaller if the precision is made greater, but this has an adverse effect on performance. For more details see Appendix~\ref{sec:gmp}.

Next, one needs to perform the remaining sums over momenta, numerically. For smaller lattices it is possible to do the full summation, but otherwise we employ a flat-weight Monte Carlo (see Appendix \ref{sec:benchmark}; for alterinative algorithm usuful in the case of local self-energy, see Appendix~\ref{sec:local_mc}). In each step, we select randomly the internal momenta $\mathbf{k}_1..\mathbf{k}_M$, and evaluate all $A_\kappa$, and permanently store the triplets $(\omega_\kappa, m_\kappa, A_\kappa)$. We perform ``on the fly'' integration for any reappearing values of $\omega_\kappa$. Even for modest lattice sizes, the number of possible values of $\omega_\kappa$ will be very large. 
To avoid immense outputs, we project $\varepsilon(\mathbf{k})$ on a uniform energy grid, so that linear combinations of $\varepsilon$'s and thus  $\omega_\kappa$'s always fall on the same uniform grid. 
The small shifts discussed in the previous paragraph also fall on a uniform grid of a much smaller step, so there will generally be several values of $\omega_\kappa$ concentrated around each point in the ``big'' $\varepsilon$-grid. This way, the number of different values of $\omega_\kappa$ one can obtain is determined by the resolution of the energy grid, i.e. the step  $\Delta\varepsilon$. Again, this is a well controlled approximation, and one can easily push the resolution so that the approximation is negligible compared to statistical noise. See Appendix \ref{sec:epsk_grid} for details.

Note also that it is essential for performance to store the different values of $\omega_\kappa,\omega_{\zeta\varsigma},\omega_{\varrho\varsigma}$ and the corresponding $\partial^r n_\eta(\omega)$, and reuse them whenever possible during the Monte Carlo sampling.

The Monte Carlo run is then performed for a given choice of the external momentum, temperature, lattice geometry and the Hartree-shifted chemical potential $\mu-Un_\sigma$ (the doping can be determined a posteriori).
Once enough measurements of $(\omega_\kappa, m_\kappa, A_\kappa)$ have been collected,
the result for $\Sigma_\mathbf{k}(z)$ for any $z$ and any $U$ can be obtained using Eq.~\ref{eq:sum_of_poles} and then Eq.~\ref{eq:total} (with $i\omega\rightarrow z$).
However, the result is a discrete set of poles on the real axis,
and requires regularization, similarly as in exact diagonalization techniques.
If it were just the simple poles on a dense uniform energy grid with a step $\Delta\varepsilon$, one could easily interpret $\mathrm{Im}\Sigma_\mathbf{k}(\omega+i0^+)$ as continuous, but known with a finite resolution, simply through $\mathrm{Im}\Sigma_\mathbf{k}(\omega_\kappa+i0^+) = -\pi A_\kappa/\Delta\varepsilon$. 
An analogous scheme could be performed even for higher-order poles on a uniform grid, order by order\cite{comment}.

The problem is that the poles are not only on a uniform grid, but rather cluster around the grid points, due to the small numerical shifts discussed previously. 
It is also impossible to separate poles according to their order because multiple poles can combine to effectively form a single higher-order pole.
This makes it very difficult to construct a binning scheme that would reinterpret the result directly on the real axis.
A better strategy is to use broadening, i.e. evaluate the self-energy slightly away from the real axis, $\Sigma(\omega+i\eta)$.
In our calculation, statistical noise dominates close to the real-axis, 
thus we take $\eta$ just large enough so that $\Sigma(\omega+i\eta)$ is a smooth function of $\omega$.

To recover the desired $\omega+i0^+$ result, one can perform a fit based on the obtained $\Sigma(\omega+i\eta)$ and the Hilbert transform 
\begin{equation}\label{eq:hilbert_def}
 \Sigma(\omega+i\eta) = -\frac{1}{\pi}\int d\varepsilon\frac{\mathrm{Im}\Sigma^\mathrm{fit}(\varepsilon)}{(\omega+i\eta)-\varepsilon}
\end{equation}
This procedure becomes trivial with $\eta\rightarrow 0$, it treats all frequencies on equal footing and is much better defined than $\Sigma(i\omega_n)\rightarrow\Sigma(\omega+i0^+)$ whenever $\eta$ is small.
Let us emphasize that the only limitation in taking a small $\eta$ is the numerical noise: when the statistical error bars are small, the procedure is very reliable,
numerically stable and does not require additional input (such as, e.g., the default model for MEM). This is
illustrated in the Appendix \ref{sec:benchmark}, where the algorithm is benchmarked against the numerical renormalization group (NRG)\cite{ZitkoPRB2009} for the solution of an Anderson impurity model\cite{hewson1993}.


\section{Results}
We have benchmarked our method carefully on several simple examples (see Appendix \ref{sec:benchmark}).
We now consider a $32\times 32$ cyclic Hubbard lattice at temperature $T=0.1D$ and $\mu-Un_\sigma=-0.1D$ (hole doping).
In this case we benchmark our method against 8-th order $\Sigma$Det~\cite{moutenet2018,fedor_2017} in imaginary frequency and find excellent agreement (see Appendix \ref{sec:bench_32x32}).

In Fig.~\ref{fig:32x32w} we show the results for
$\mathrm{Im}\Sigma(\omega+i\eta)$ close to the real axis (finite $\eta<\pi T$,
lower than the first fermionic Matsubara frequency). Closer than this, stronger
noisy features start to appear. Let us emphasize that the
statistical noise is far more pronounced on the real axis, i.e. convergence on
the imaginary axis does not necessarily imply convergence on the real axis.
Different lines represent calculations with different maximal perturbation orders $N_\mathrm{max}$, at 6 characteristic $\mathbf{k}$-points and 2 values of $U$. 
The shaded region is a piecewise-trapezoid $\mathrm{Im}\Sigma^\mathrm{fit}(\omega+i0^+)$ obtained with resolution $\Delta\omega = 1.6\eta$.

At $U=1D$ fifth order diagrams contribute very little and the result is
practically converged with respect to $N_\mathrm{max}$.
%
At $U=1.5D$, the result is not fully converged by order 5, but is apparently
close to convergence.
We observe several non-causal features
$\mathrm{Im}\Sigma_\mathbf{k}(\omega)>0$. At large negative $\omega$, this
happens at $\mathbf{k}=(0,0)$ at order 4, but is then fixed by order 5. At
large positive $\omega$, the problem appears at order 5, and is likely to be
fixed by higher orders in perturbation. These non-casual features do not appear
to be artifacts of the statistical noise but rather a result of the truncation of
the perturbation series. This calls for great caution in the use of MEM.
Indeed, MEM performed with built-in causality is bound to miss any such
features and may compensate for them in an uncontrolled way. 

It is interesting that in most cases $\mathrm{Im}\Sigma(\omega)$ features two
broad peaks with a dip around $\omega=0$. However, at $U=1.5D$ around
$\mathbf{k}=(0,\pi)$, a third peak appears close to $\omega=0$. We interpret this
peak as a precursor for the pseudogap behavior: as temperature further
decreases at this doping (around $5\%$), the peak may approach $\omega=0$ and
induce a larger, insulating-like self-energy as observed in imaginary-time
calculations, e.g. Ref.~\onlinecite{wuPRB2017}.

Finally, the panels on the right present the filled part of the corresponding $\mathbf{k}$-resolved spectral functions.
These plots are relevant for recent spectral function measurements in optical lattice realizations of the Hubbard model\cite{brown2019}.
One can observe that the spectral function preserves the general form of the non-interacting limit, but spans a bigger energy range and becomes more incoherent (wider lines of lesser intensity) as interaction is increased.

\section{Conclusions and prospects}
We have resolved the main conceptual issues regarding the application of algorithmic Matsubara summations in the context of diagrammatic Monte Carlo.
This includes the precision and efficiency concerns in the evaluation of the pole amplitudes, as well as the extraction of the real-axis results.
There is possibility for further optimization which will likely allow to push the method to higher perturbation orders in the future.


We demonstrate that our method is readily useful in the study of the single-particle spectra in the intermediate coupling regime of the Hubbard model, which has been the subject of recent publications~\cite{simkovic2018,kim2019,brown2019}.
Finally, our method holds great promise for future work in the cases where analytical continuation is particularly difficult.
These include, for example, the high-temperature and calculations of the current-current correlation function $\Lambda(\omega)$\cite{vucicevicPRL2019}.
Our approach even allows for a straightforward restriction to a selected window of energies - if one is interested in dc resistivity, one may calculate $\Lambda(\omega)$ only at very low frequency and that way gain an important speedup.

\begin{acknowledgments}
We are grateful to Rok \v Zitko for providing NRG data. The exact-diagonalization
results were obtained using the PyED code, written by Hugo Strand\cite{strandPyED}. The continuous-time
interaction expansion algorithm\cite{rubtsov2004,rubtsov_prb_2005} was developed using the
TRIQS~\cite{triqs} library. Computations were
performed on the PARADOX supercomputing facility (Scientific
Computing Laboratory of the Institute of Physics Belgrade) and ALPHA cluster (Coll\`ege de France) as well as using HPC resources from
GENCI (Grant No. A0050510609). We thank Mihailo \v Cubrovi\'c for his help
with the preparation of the manuscript.
J.~V is supported by the Serbian Ministry of Education,
Science and Technological Development under Project No. ON171017.
\end{acknowledgments}

\appendix

\section{Formalism details} \label{sec:formalism}

\subsection{Derivation of Eq.~\ref{eq:main_transformation}} \label{sec:eq6}

The partial fraction expansion employs the residue theorem, and the textbook expression reads
\begin{eqnarray}\label{eq:partial_expansion}
&&\prod_\gamma \frac{1}{(z-z_\gamma)^{m_\gamma}} = 
  \sum_\gamma \sum_{r=1}^{m_\gamma} \frac{1}{(z-z_\gamma)^{r}} \times \\ \nonumber
&& \;\;\;\;\;\;\;\;\;\;\times\frac{1}{(m_\gamma - r)!}  \lim_{z\rightarrow z_\gamma} \partial^{m_\gamma-r}_z
            \prod_{\gamma'\neq \gamma} \frac{1}{(z-z_{\gamma'})^{m_{\gamma'}}}
\end{eqnarray}

The derivative of a product of poles can be expressed in the following way
\begin{eqnarray} \nonumber
 && \partial^{n}_z \prod_{\gamma} \frac{1}{(z-z_\gamma)^{m_\gamma}} = (-1)^n  n! \sum_{{\cal C}\{p_\gamma \in {\mathbb N_0}\}:\sum_\gamma p_\gamma = n} \times \\ \label{eq:derivative}
 && \;\;\;\;\;\;\;\;\;\times \prod_\gamma\frac{(m_\gamma+p_\gamma-1)!}{p_\gamma!(m_\gamma-1)!} \frac{1}{(z-z_\gamma)^{m_\gamma+p_\gamma}}
\end{eqnarray}
Here the sum goes over all combinations $\cal C$ of a choice of a non-negative-integer $p$  per pole $\gamma$, such that their sum is $n$.

Putting together the equations Eq.~\ref{eq:partial_expansion} and Eq.~\ref{eq:derivative}, one obtains Eq.~\ref{eq:main_transformation}.

The derivation of Eq.\ref{eq:derivative} relies on performing 
$ \partial_z [f(z)g(z)] = [\partial_z f(z)] g(z)+ f(z)[\partial_z g(z)]$
and 
$\partial_z\frac{1}{(z-z_\gamma)^{m_\gamma}}=-m_\gamma\frac{1}{(z-z_\gamma)^{m_\gamma+1}}$, recursively.
Having these in mind, it is clear that the final result will consist of a number of terms, each term being a product of the original poles, some with increased orders.
In each term, we will have acted with the derivative upon each pole $\gamma$ a certain number of times $p_\gamma \geq 0$, so as to use up all the derivatives, i.e. $\sum_\gamma p_\gamma=n$. 
For each pole that is acted upon at least once, this leads to $\partial^{p_\gamma}_z\frac{1}{(z-z_\gamma)^{m_\gamma}}=(-1)^{p_\gamma} m_\gamma(m_\gamma+1)...(m_\gamma+p_\gamma-1)\frac{1}{(z-z_\gamma)^{m_\gamma+p_\gamma}}$. Hence the overall sign $\prod_\gamma (-1)^{p_\gamma}=(-1)^n$.
However, we can apply derivatives in any order, so there is also a combinatorial factor corresponding to permutation of multisets $n!/(\prod_\gamma p_\gamma!)$ (number of distinct anagrams of an $n$-long word consisting of unique letters indexed by $\gamma$, each appearing $p_\gamma$ times in the word).

Let's check and illustrate Eq.\ref{eq:derivative} on a simple example, where one can carry out the derivatives by hand. Say
\begin{eqnarray}\label{eq:example}
 &&\partial^3_z \frac{1}{z-z_1}\frac{1}{(z-z_2)^2} \\ \nonumber
  &&= -6 \Bigg(4\frac{1}{z-z_1}\frac{1}{(z-z_2)^5} +3\frac{1}{(z-z_1)^2}\frac{1}{(z-z_2)^4}  \\  \nonumber
 && \;\;\;\;+2\frac{1}{(z-z_1)^3}\frac{1}{(z-z_2)^3} + \frac{1}{(z-z_1)^4}\frac{1}{(z-z_2)^2} \Bigg)
\end{eqnarray}

We can immediately identify the prefactor $(-1)^n n! = (-1)^3 3! = -6$. Also, we see there are 4 terms corresponding to 4 possible choices of $(p_1,p_2)$ such that $p_1+p_2=n=3$, respectively
\begin{equation}
 {\cal C} = \{(0,3),(1,2),(2,1),(3,0)\}
\end{equation}
Now the prefactors $\prod_\gamma (m_\gamma+p_\gamma-1)!/(p_\gamma!(m_\gamma-1)!)$ can be evaluated for each combination
\begin{eqnarray}
 (0,3): \frac{(1+0-1)!}{0!0!}\frac{(2+3-1)!}{3!1!} = \frac{1}{1}\frac{4!}{3!} = 4 \\ \nonumber
 (1,2): \frac{(1+1-1)!}{1!0!}\frac{(2+2-1)!}{2!1!} = \frac{1}{1}\frac{3!}{2!} = 3 \\ \nonumber
 (2,1): \frac{(1+2-1)!}{2!0!}\frac{(2+1-1)!}{1!1!} = \frac{2!}{2!}\frac{2!}{1} = 2 \\ \nonumber
 (3,0): \frac{(1+3-1)!}{3!0!}\frac{(2+0-1)!}{0!1!} = \frac{3!}{3!}\frac{1}{1} = 1 
\end{eqnarray}
all of which we can readily identify on the right-hand side of Eq.\ref{eq:example}.

\subsection{Numbers of poles and terms per diagram} \label{sec:poles_and_terms_stats}

The Eq.~\ref{eq:sum_of_poles} in the main text is the final result of Matsubara summations for a given diagram. 
It is a sum of a a certain number $N_\mathrm{poles}$ of distinct poles $(\omega_\kappa,m_\kappa)$,
each with $N_\mathrm{terms}$ distinct terms in the amplitude $A_\kappa$.
We tabulate in Table~\ref{table:stats} the range and the geometrical average (typical value) of these numbers for each perturbation order $N$.

\begin{table}
  \begin{tabular}{  l|cccc }
    $N$ & $N_\mathrm{poles}$ & $N^\mathrm{typ}_\mathrm{poles}$ & $N_\mathrm{terms}$  & $N^\mathrm{typ}_\mathrm{terms}$   \\ \hline
    2 & 1 & 1 & 4 & 4 \\
    3 & 2 & 2 & 12-14 & 13\\
    4 & 3-4 & 3.5 & 16-70 & 29.7\\
    5 & 4-8 & 5.6 & 32-482 & 97.9\\
    6 & 5-14 & 8.9 & 32-5092 & 296.2 
  \end{tabular}
\caption{Numbers of poles and terms in the symbolic expression obtained by analytical Matsubara summations.}
\label{table:stats}
\end{table}
\newpage
\begin{figure}
 \includegraphics[width=1.9in]{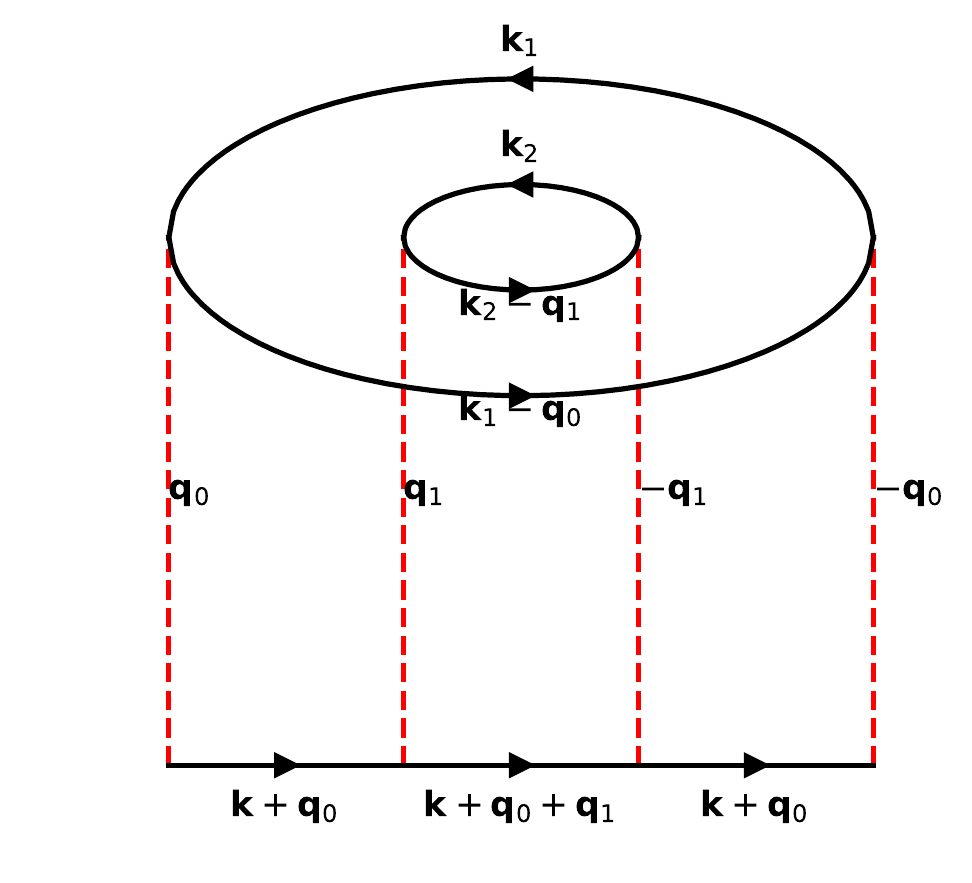}
 \caption{An example of a momentum-labeled 4th order diagram on the lattice.
 }
 \label{fig:diagramexample}
\end{figure}

\begin{widetext}

\subsection{Results of symbolic algebra} \label{sec:eq_example}
We present here an example of the analytic expression for the contribution of a self-energy diagram. We choose the 4th order diagram presented in Fig.~\ref{fig:diagramexample}.
We start from the expression of the form Eq.~\ref{eq:diag_contrib}
\begin{eqnarray}
 &&{\cal D}_\mathbf{k}(i\omega) = (-1)^{2}\sum_{\mathbf{k}_1,\mathbf{k}_2}\sum_{\mathbf{q}_0,\mathbf{q}_1} \sum_{i\omega_1,i\omega_2}\sum_{i\nu_0,i\nu_1} \\ \nonumber
 &&\;\;\;\;\;\; 
 G^\mathrm{HF}_{\mathbf{k}+\mathbf{q}_0}(i\omega+i\nu_0)
 G^\mathrm{HF}_{\mathbf{k}+\mathbf{q}_0+\mathbf{q}_1}(i\omega+i\nu_0+i\nu_1)
 G^\mathrm{HF}_{\mathbf{k}+\mathbf{q}_0}(i\omega+i\nu_0)
 G^\mathrm{HF}_{\mathbf{k}_1}(i\omega_1)
 G^\mathrm{HF}_{\mathbf{k}_1-\mathbf{q}_0}(i\omega_1-i\nu_0)
 G^\mathrm{HF}_{\mathbf{k}_2}(i\omega_2)
 G^\mathrm{HF}_{\mathbf{k}_2-\mathbf{q}_1}(i\omega_2-i\nu_1) \\ \nonumber
 &&\;\;\;\; = (-1)^{2}\sum_{\mathbf{k}_1,\mathbf{k}_2}\sum_{\mathbf{q}_0,\mathbf{q}_1} \sum_{i\omega_1,i\omega_2}\sum_{i\nu_0,i\nu_1} \\ \nonumber
 &&\;\;\;\;\;\; 
 \left( \frac{1}{i\omega+i\nu_0 - \varepsilon_{\mathbf{k}+\mathbf{q}_0}} \right)^2
 \frac{1}{i\omega+i\nu_0+i\nu_1-\varepsilon_{\mathbf{k}+\mathbf{q}_0+\mathbf{q}_1}}
 \frac{1}{i\omega_1-\varepsilon_{\mathbf{k}_1}}
 \frac{1}{i\omega_1-i\nu_0-\varepsilon_{\mathbf{k}_1-\mathbf{q}_0}}
 \frac{1}{i\omega_2-\varepsilon_{\mathbf{k}_2}}
 \frac{1}{i\omega_2-i\nu_1-\varepsilon_{\mathbf{k}_2-\mathbf{q}_1}}
\end{eqnarray}
Here we have already imposed momentum-conservation, which leaves only two internal bosonic frequencies/momenta to be summed over (independent momenta carried by fermions and vertices are denoted in Fig.~\ref{fig:diagramexample}). For the sake of notational brevity, here, as well as in the rest of the Appendix, we take $\varepsilon_\mathbf{k}\equiv \varepsilon(\mathbf{k})$.

The first step in performing the analytical Matsubara frequency summations is to choose one internal Matsubara frequency, and then isolate the factors (poles) which depend on it. Say, we choose $i\nu_0$. We can regroup the factors conveniently
\begin{eqnarray}
 &&{\cal D}_\mathbf{k}(i\omega) = (-1)^{2}\sum_{\mathbf{k}_1,\mathbf{k}_2}\sum_{\mathbf{q}_0,\mathbf{q}_1} \sum_{i\omega_1,i\omega_2}\sum_{i\nu_1} \\ \nonumber
 &&\;\;\;\;\;\; 
 \frac{1}{i\omega_1-\varepsilon_{\mathbf{k}_1}}
 \frac{1}{i\omega_2-\varepsilon_{\mathbf{k}_2}}
 \frac{1}{i\omega_2-i\nu_1-\varepsilon_{\mathbf{k}_2-\mathbf{q}_1}}
 \sum_{i\nu_0}
 \frac{1}{\left(i\nu_0 -(-i\omega + \varepsilon_{\mathbf{k}+\mathbf{q}_0}) \right)^2}
 \frac{1}{i\nu_0 -(-i\omega-i\nu_1+\varepsilon_{\mathbf{k}+\mathbf{q}_0+\mathbf{q}_1})}
 \frac{-1}{i\nu_0-(i\omega_1-\varepsilon_{\mathbf{k}_1-\mathbf{q}_0})}
\end{eqnarray}
Now the expression has the form of a product of poles with respect to $i\nu_0$ (Eq.~\ref{eq:product_of_poles}), where the rest can be considered a prefactor (denoted $P$).
The product of poles can be then transformed using the main transformation Eq.~\ref{eq:main_transformation}
\begin{eqnarray}
 &&{\cal D}_\mathbf{k}(i\omega) = (-1)^{2}\sum_{\mathbf{k}_1,\mathbf{k}_2}\sum_{\mathbf{q}_0,\mathbf{q}_1} \sum_{i\omega_1,i\omega_2}\sum_{i\nu_1}
  \frac{1}{i\omega_1-\varepsilon_{\mathbf{k}_1}}
 \frac{1}{i\omega_2-\varepsilon_{\mathbf{k}_2}}
 \frac{1}{i\omega_2-i\nu_1-\varepsilon_{\mathbf{k}_2-\mathbf{q}_1}}
 \times\\ \nonumber
 && \;\;\;\;\;
 \Bigg\{
 \frac{1}{\left(i\omega_1-\varepsilon_{\mathbf{k}_1-\mathbf{q}_0} -(-i\omega + \varepsilon_{\mathbf{k}+\mathbf{q}_0}) \right)^2}
 \frac{1}{i\omega_1-\varepsilon_{\mathbf{k}_1-\mathbf{q}_0}-(-i\omega-i\nu_1+\varepsilon_{\mathbf{k}+\mathbf{q}_0+\mathbf{q}_1})}
 \sum_{i\nu_0}\frac{-1}{i\nu_0-(i\omega_1-\varepsilon_{\mathbf{k}_1-\mathbf{q}_0})} \\ \nonumber
 && \;\;\;\;\;\;\;\;+ 
 \frac{1}{\left(-i\omega-i\nu_1+\varepsilon_{\mathbf{k}+\mathbf{q}_0+\mathbf{q}_1} -(-i\omega + \varepsilon_{\mathbf{k}+\mathbf{q}_0}) \right)^2}
 \frac{-1}{-i\omega-i\nu_1+\varepsilon_{\mathbf{k}+\mathbf{q}_0+\mathbf{q}_1}-(i\omega_1-\varepsilon_{\mathbf{k}_1-\mathbf{q}_0})}
 \sum_{i\nu_0}\frac{1}{i\nu_0 -(-i\omega-i\nu_1+\varepsilon_{\mathbf{k}+\mathbf{q}_0+\mathbf{q}_1})} \\ \nonumber
 && \;\;\;\;\;\;\;\;-
 \Bigg[
   \left(\frac{1}{-i\omega + \varepsilon_{\mathbf{k}+\mathbf{q}_0} -(-i\omega-i\nu_1+\varepsilon_{\mathbf{k}+\mathbf{q}_0+\mathbf{q}_1})}\right)^2
   \frac{-1}{-i\omega + \varepsilon_{\mathbf{k}+\mathbf{q}_0}-(i\omega_1-\varepsilon_{\mathbf{k}_1-\mathbf{q}_0})} \\ \nonumber
 && \;\;\;\;\;\;\;\;\;\;\;\;+
   \frac{1}{-i\omega + \varepsilon_{\mathbf{k}+\mathbf{q}_0} -(-i\omega-i\nu_1+\varepsilon_{\mathbf{k}+\mathbf{q}_0+\mathbf{q}_1})}
   \left(\frac{-1}{-i\omega + \varepsilon_{\mathbf{k}+\mathbf{q}_0}-(i\omega_1-\varepsilon_{\mathbf{k}_1-\mathbf{q}_0})}\right)^2 
 \Bigg]
 \sum_{i\nu_0}\frac{1}{i\nu_0 -(-i\omega + \varepsilon_{\mathbf{k}+\mathbf{q}_0})} \\ \nonumber
 && \;\;\;\;\;\;\;\;+
 \frac{1}{-i\omega + \varepsilon_{\mathbf{k}+\mathbf{q}_0} -(-i\omega-i\nu_1+\varepsilon_{\mathbf{k}+\mathbf{q}_0+\mathbf{q}_1})}
 \frac{-1}{-i\omega + \varepsilon_{\mathbf{k}+\mathbf{q}_0}-(i\omega_1-\varepsilon_{\mathbf{k}_1-\mathbf{q}_0})}
 \sum_{i\nu_0}\frac{1}{\left(i\nu_0 -(-i\omega + \varepsilon_{\mathbf{k}+\mathbf{q}_0}) \right)^2}
 \Bigg\}
\end{eqnarray}
We can now evaluate the Matsubara frequency summations per $i\nu_0$, using Eq.~\ref{eq:mats_sum} and then Eq.~\ref{eq:remove_imaginary_in_distrib_arg}. Then, the denominators can be simplified at the symbolic level.
\begin{eqnarray}
 &&{\cal D}_\mathbf{k}(i\omega) = (-1)^{2}\sum_{\mathbf{k}_1,\mathbf{k}_2}\sum_{\mathbf{q}_0,\mathbf{q}_1} \sum_{i\omega_1,i\omega_2}\sum_{i\nu_1} 
  \frac{1}{i\omega_1-\varepsilon_{\mathbf{k}_1}}
 \frac{1}{i\omega_2-\varepsilon_{\mathbf{k}_2}}
 \frac{1}{i\omega_2-i\nu_1-\varepsilon_{\mathbf{k}_2-\mathbf{q}_1}}
 \times\\ \nonumber
 && \;\;\;\;\;
 \Bigg\{
 \frac{1}{\left(i\omega+i\omega_1-\varepsilon_{\mathbf{k}_1-\mathbf{q}_0} - \varepsilon_{\mathbf{k}+\mathbf{q}_0} \right)^2}
 \frac{1}{i\omega+i\nu_1+i\omega_1-\varepsilon_{\mathbf{k}_1-\mathbf{q}_0}-\varepsilon_{\mathbf{k}+\mathbf{q}_0+\mathbf{q}_1}}
 (-)^3n_F(-\varepsilon_{\mathbf{k}_1-\mathbf{q}_0}) \\ \nonumber
 && \;\;\;\;\;\;\;\;+ 
 \frac{1}{\left(-i\nu_1+\varepsilon_{\mathbf{k}+\mathbf{q}_0+\mathbf{q}_1} - \varepsilon_{\mathbf{k}+\mathbf{q}_0} \right)^2}
 \frac{-1}{-i\omega-i\omega_1-i\nu_1+\varepsilon_{\mathbf{k}+\mathbf{q}_0+\mathbf{q}_1}+\varepsilon_{\mathbf{k}_1-\mathbf{q}_0}}
 (-)^2n_F(\varepsilon_{\mathbf{k}+\mathbf{q}_0+\mathbf{q}_1}) \\ \nonumber
 && \;\;\;\;\;\;\;\;-
 \Bigg[
   \left(\frac{1}{i\nu_1 + \varepsilon_{\mathbf{k}+\mathbf{q}_0} -\varepsilon_{\mathbf{k}+\mathbf{q}_0+\mathbf{q}_1}}\right)^2
   \frac{-1}{-i\omega -i\omega_1 + \varepsilon_{\mathbf{k}+\mathbf{q}_0}+\varepsilon_{\mathbf{k}_1-\mathbf{q}_0}} \\ \nonumber
 && \;\;\;\;\;\;\;\;\;\;\;\;+
   \frac{1}{i\nu_1 + \varepsilon_{\mathbf{k}+\mathbf{q}_0} -\varepsilon_{\mathbf{k}+\mathbf{q}_0+\mathbf{q}_1}}
   \left(\frac{-1}{-i\omega-i\omega_1 + \varepsilon_{\mathbf{k}+\mathbf{q}_0}+\varepsilon_{\mathbf{k}_1-\mathbf{q}_0}}\right)^2 
 \Bigg]
 (-)^2n_F(\varepsilon_{\mathbf{k}+\mathbf{q}_0}) \\ \nonumber
 && \;\;\;\;\;\;\;\;+
 \frac{1}{i\nu_1 + \varepsilon_{\mathbf{k}+\mathbf{q}_0} -\varepsilon_{\mathbf{k}+\mathbf{q}_0+\mathbf{q}_1}}
 \frac{-1}{-i\omega -i\omega_1+ \varepsilon_{\mathbf{k}+\mathbf{q}_0}+\varepsilon_{\mathbf{k}_1-\mathbf{q}_0}}
 (-)^2\partial n_F(\varepsilon_{\mathbf{k}+\mathbf{q}_0} )
 \Bigg\}
\end{eqnarray}
The procedure can now be repeated for the next choice of the Matsubara variable.

We now present the final result of the symbolic algorithm for the diagram presented in Fig.~\ref{fig:diagramexample}.
The diagram contributes one second-order pole and two simple poles. The number of terms in the amplitudes for each pole is 16, 24 and 16, respectively. For the expression to fit a single page, we only show several representative terms in the amplitude of each pole. 
\pagebreak
\begin{eqnarray} \label{eq:symres_example}
{\cal D}_\mathbf{k}(z) &=& (-1)^{2}\sum_{\mathbf{k}_1,\mathbf{k}_2}\sum_{\mathbf{q}_0,\mathbf{q}_1} \times \Bigg\{ \\ \nonumber
\nonumber && \frac{1}{ ( z + \varepsilon_{\mathbf{k}_1} - \varepsilon_{\mathbf{k}_1-\mathbf{q}_0} - \varepsilon_{\mathbf{k}+\mathbf{q}_0} )^2 } \Bigg[ \\ 
\nonumber && \;\;\;\; n_F\left( \varepsilon_{\mathbf{k}_2-\mathbf{q}_1} \right) n_F\left( \varepsilon_{\mathbf{k}_1-\mathbf{q}_0} \right) n_F\left( \varepsilon_{\mathbf{k}+\mathbf{q}_0+\mathbf{q}_1} \right) \frac{1}{ \varepsilon_{\mathbf{k}_2} - \varepsilon_{\mathbf{k}+\mathbf{q}_0+\mathbf{q}_1} - \varepsilon_{\mathbf{k}_2-\mathbf{q}_1} + \varepsilon_{\mathbf{k}+\mathbf{q}_0} } n_F\left( \varepsilon_{\mathbf{k}+\mathbf{q}_0} \right) \\
\nonumber  && \;\;\;\;+ n_F\left( \varepsilon_{\mathbf{k}_2-\mathbf{q}_1} \right) n_F\left( \varepsilon_{\mathbf{k}_1-\mathbf{q}_0} \right) n_B\left( \varepsilon_{\mathbf{k}_2} - \varepsilon_{\mathbf{k}_2-\mathbf{q}_1} \right) \frac{1}{ \varepsilon_{\mathbf{k}_2} - \varepsilon_{\mathbf{k}+\mathbf{q}_0+\mathbf{q}_1} - \varepsilon_{\mathbf{k}_2-\mathbf{q}_1} + \varepsilon_{\mathbf{k}+\mathbf{q}_0} } n_F\left( \varepsilon_{\mathbf{k}+\mathbf{q}_0} \right) \\
\nonumber  && \;\;\;\;- n_F\left( \varepsilon_{\mathbf{k}_2-\mathbf{q}_1} \right) n_F\left( \varepsilon_{\mathbf{k}_1} \right) n_F\left( \varepsilon_{\mathbf{k}+\mathbf{q}_0+\mathbf{q}_1} \right) \frac{1}{ \varepsilon_{\mathbf{k}_2} - \varepsilon_{\mathbf{k}+\mathbf{q}_0+\mathbf{q}_1} - \varepsilon_{\mathbf{k}_2-\mathbf{q}_1} + \varepsilon_{\mathbf{k}+\mathbf{q}_0} } n_F\left( \varepsilon_{\mathbf{k}+\mathbf{q}_0} \right) \\
\nonumber  && \;\;\;\;- n_F\left( \varepsilon_{\mathbf{k}_2} \right) n_F\left( \varepsilon_{\mathbf{k}_1-\mathbf{q}_0} \right) n_B\left( \varepsilon_{\mathbf{k}_2} - \varepsilon_{\mathbf{k}_2-\mathbf{q}_1} \right) \frac{1}{ \varepsilon_{\mathbf{k}_2} - \varepsilon_{\mathbf{k}+\mathbf{q}_0+\mathbf{q}_1} - \varepsilon_{\mathbf{k}_2-\mathbf{q}_1} + \varepsilon_{\mathbf{k}+\mathbf{q}_0} } n_F\left( \varepsilon_{\mathbf{k}+\mathbf{q}_0} \right) \\
\nonumber  && \;\;\;\; + ...\\
\nonumber  && \Bigg] \\
\nonumber  && + \frac{1}{ z + \varepsilon_{\mathbf{k}_1} - \varepsilon_{\mathbf{k}_1-\mathbf{q}_0} - \varepsilon_{\mathbf{k}+\mathbf{q}_0} } \Bigg[ \\ 
\nonumber && \;\;\;\; n_F\left( \varepsilon_{\mathbf{k}_2-\mathbf{q}_1} \right) n_F\left( \varepsilon_{\mathbf{k}_1-\mathbf{q}_0} \right) n_F\left( \varepsilon_{\mathbf{k}+\mathbf{q}_0+\mathbf{q}_1} \right) \frac{1}{ \varepsilon_{\mathbf{k}_2} - \varepsilon_{\mathbf{k}+\mathbf{q}_0+\mathbf{q}_1} - \varepsilon_{\mathbf{k}_2-\mathbf{q}_1} + \varepsilon_{\mathbf{k}+\mathbf{q}_0} } \partial n_F\left( \varepsilon_{\mathbf{k}+\mathbf{q}_0} \right) \\
\nonumber  && \;\;\;\;+ n_F\left( \varepsilon_{\mathbf{k}_2-\mathbf{q}_1} \right) n_F\left( \varepsilon_{\mathbf{k}_1} \right) n_F\left( \varepsilon_{\mathbf{k}+\mathbf{q}_0+\mathbf{q}_1} \right) n_B\left( \varepsilon_{\mathbf{k}_1} - \varepsilon_{\mathbf{k}_1-\mathbf{q}_0} \right) \frac{1}{ ( \varepsilon_{\mathbf{k}_2} + \varepsilon_{\mathbf{k}+\mathbf{q}_0} - \varepsilon_{\mathbf{k}+\mathbf{q}_0+\mathbf{q}_1} - \varepsilon_{\mathbf{k}_2-\mathbf{q}_1} )^2 } \\
\nonumber  && \;\;\;\;- n_F\left( \varepsilon_{\mathbf{k}_2} \right) n_F\left( \varepsilon_{\mathbf{k}_1-\mathbf{q}_0} \right) n_F\left( \varepsilon_{\mathbf{k}+\mathbf{q}_0+\mathbf{q}_1} \right) \frac{1}{ \varepsilon_{\mathbf{k}_2} - \varepsilon_{\mathbf{k}+\mathbf{q}_0+\mathbf{q}_1} - \varepsilon_{\mathbf{k}_2-\mathbf{q}_1} + \varepsilon_{\mathbf{k}+\mathbf{q}_0} } \partial n_F\left( \varepsilon_{\mathbf{k}+\mathbf{q}_0} \right) \\
\nonumber  && \;\;\;\;- n_F\left( \varepsilon_{\mathbf{k}_2} \right) n_F\left( \varepsilon_{\mathbf{k}_1} \right) n_B\left( \varepsilon_{\mathbf{k}_2} - \varepsilon_{\mathbf{k}_2-\mathbf{q}_1} \right) \frac{1}{ ( \varepsilon_{\mathbf{k}_2} - \varepsilon_{\mathbf{k}+\mathbf{q}_0+\mathbf{q}_1} - \varepsilon_{\mathbf{k}_2-\mathbf{q}_1} + \varepsilon_{\mathbf{k}+\mathbf{q}_0} )^2 } n_F\left( \varepsilon_{\mathbf{k}+\mathbf{q}_0} \right) \\
\nonumber  && \;\;\;\;+ ... \\
\nonumber  &&\Bigg] \\
\nonumber  && + \frac{1}{ z + \varepsilon_{\mathbf{k}_2} + \varepsilon_{\mathbf{k}_1} - \varepsilon_{\mathbf{k}+\mathbf{q}_0+\mathbf{q}_1} - \varepsilon_{\mathbf{k}_2-\mathbf{q}_1} - \varepsilon_{\mathbf{k}_1-\mathbf{q}_0} } \Bigg[ \\ 
\nonumber && \;\;\;\; n_F\left( \varepsilon_{\mathbf{k}_2-\mathbf{q}_1} \right) n_F\left( \varepsilon_{\mathbf{k}_1-\mathbf{q}_0} \right) n_F\left( \varepsilon_{\mathbf{k}+\mathbf{q}_0+\mathbf{q}_1} \right) \frac{1}{ ( - \varepsilon_{\mathbf{k}_2} + \varepsilon_{\mathbf{k}+\mathbf{q}_0+\mathbf{q}_1} + \varepsilon_{\mathbf{k}_2-\mathbf{q}_1} - \varepsilon_{\mathbf{k}+\mathbf{q}_0} )^2 } n_F\left( - \varepsilon_{\mathbf{k}_2} + \varepsilon_{\mathbf{k}+\mathbf{q}_0+\mathbf{q}_1} + \varepsilon_{\mathbf{k}_2-\mathbf{q}_1} \right) \\
\nonumber  && \;\;\;\;+ n_F\left( \varepsilon_{\mathbf{k}_2-\mathbf{q}_1} \right) n_F\left( \varepsilon_{\mathbf{k}_1-\mathbf{q}_0} \right) n_F\left( \varepsilon_{\mathbf{k}+\mathbf{q}_0+\mathbf{q}_1} \right) n_B\left( \varepsilon_{\mathbf{k}_1} - \varepsilon_{\mathbf{k}_1-\mathbf{q}_0} \right) \frac{1}{ ( - \varepsilon_{\mathbf{k}_2} - \varepsilon_{\mathbf{k}+\mathbf{q}_0} + \varepsilon_{\mathbf{k}+\mathbf{q}_0+\mathbf{q}_1} + \varepsilon_{\mathbf{k}_2-\mathbf{q}_1} )^2 } \\
\nonumber  && \;\;\;\;- n_F\left( \varepsilon_{\mathbf{k}_2} \right) n_F\left( \varepsilon_{\mathbf{k}_1-\mathbf{q}_0} \right) n_F\left( \varepsilon_{\mathbf{k}+\mathbf{q}_0+\mathbf{q}_1} \right) \frac{1}{ ( - \varepsilon_{\mathbf{k}_2} + \varepsilon_{\mathbf{k}+\mathbf{q}_0+\mathbf{q}_1} + \varepsilon_{\mathbf{k}_2-\mathbf{q}_1} - \varepsilon_{\mathbf{k}+\mathbf{q}_0} )^2 } n_F\left( - \varepsilon_{\mathbf{k}_2} + \varepsilon_{\mathbf{k}+\mathbf{q}_0+\mathbf{q}_1} + \varepsilon_{\mathbf{k}_2-\mathbf{q}_1} \right) \\
\nonumber  && \;\;\;\; ... \\
\nonumber  &&\Bigg] \\
\nonumber \Bigg\} &&
\end{eqnarray}
\end{widetext}
\clearpage
\newpage
\mbox{~}
\clearpage
\newpage

\subsection{Calculation of Fermi/Bose function derivatives} \label{sec:eq_distribs}

In the numerical evaluation of the amplitudes of the poles ($A_\kappa$, Eq.~\ref{eq:sum_of_poles} and Eq.~\ref{eq:amp_final}), we use the general expression for the derivatives of the Fermi/Bose distribution function
\begin{equation}
  \partial^r_\omega n_\eta(\omega) = -\beta^r \sum_{k=0}^{r} \frac{(-)^{k+1} f_{r,k} e^{k\beta \omega}}{(e^{\beta \omega}-\eta)^{k+1}}
\end{equation}
with $f_{r,k}\in\mathbb{N}_0$ tabulated below
\begin{center}
  \begin{tabular}{  l|ccccccc }
    $r\backslash k$ & 0 & 1 & 2 & 3 & 4 & 5 & 6  \\ \hline
    0 &1&&&&&& \\
    1 &0&1&&&&& \\
    2 &0&1&2&&&& \\
    3 &0&1&6&6&&& \\
    4 &0&1&14&36&24&& \\
    5 &0&1&30&150&240&120& \\
    6 &0&1&62&540&1560&1800&720 
  \end{tabular}
\end{center}

\begin{figure*}[ht!]
 \includegraphics[width=3.2in]{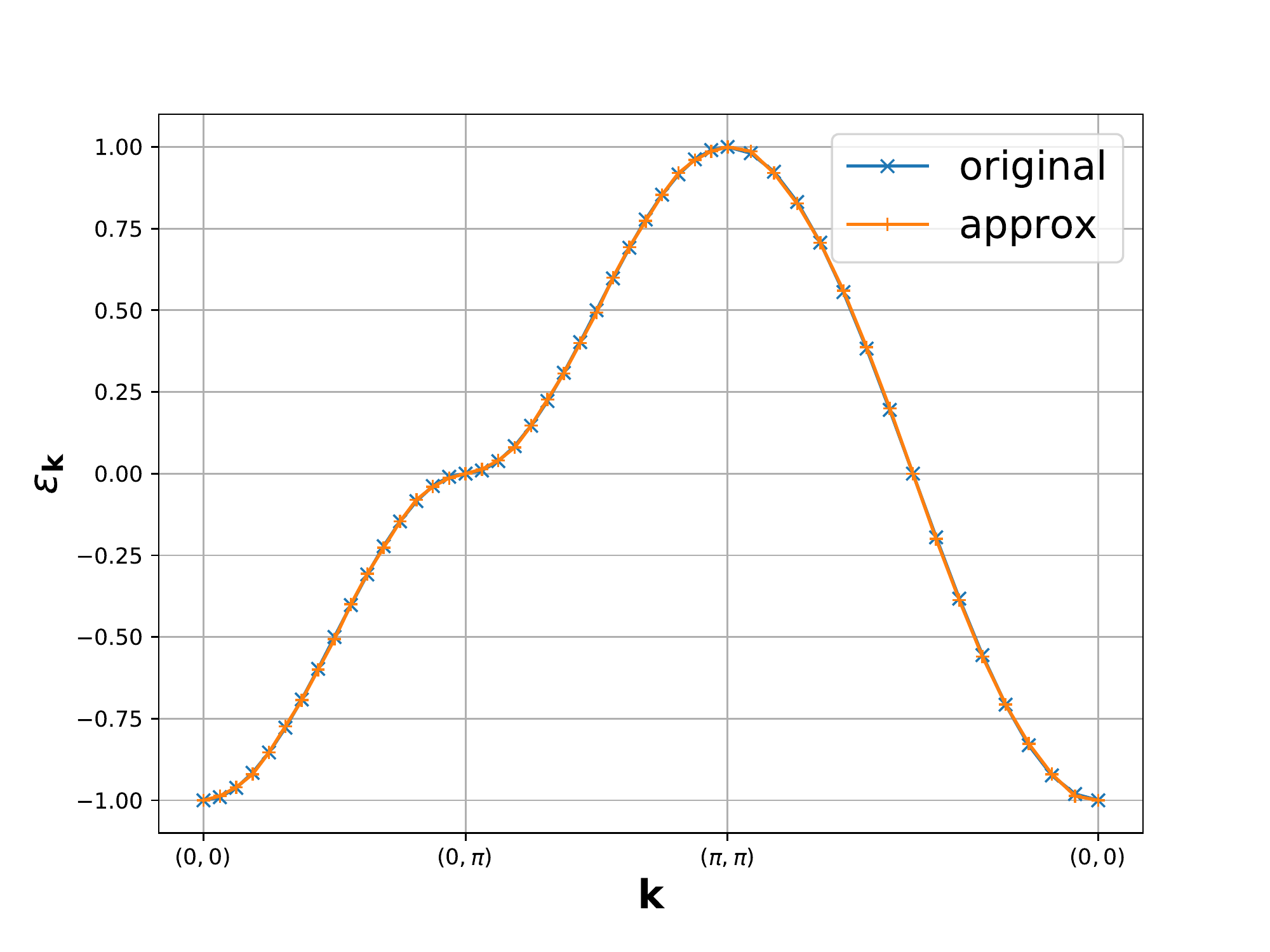}
 \caption{Approximation of the dispersion used to avoid unmanageable outputs.
 }
 \label{fig:epsk}
\end{figure*}

\subsection{Dispersion on an equidistant grid} \label{sec:epsk_grid}

We present here in detail the numerical trick that we use to avoid unmanageable outputs from the Monte Carlo summation. For a given lattice size (in our case $32\times 32$), we approximate $\varepsilon_\mathbf{k}$ so that it takes on values only from a given set $\Xi$ of equidistant numbers spanning the bandwidth (in our case the number of points is  $N_\Xi=151$). The new approximate dispersion therefore has the property
\begin{equation}
 \tilde{\varepsilon}_\mathbf{k} \in \Xi, \forall \mathbf{k}
\end{equation}
with
\begin{equation}
 \Xi = \{ \mathrm{min}_\mathbf{k} \varepsilon_\mathbf{k}+j\Delta\varepsilon \}_{j=0}^{N_\Xi-1}
\end{equation}
and 
\begin{equation} 
\Delta\varepsilon = \frac{\mathrm{max}_\mathbf{k} \varepsilon_\mathbf{k} - \mathrm{min}_\mathbf{k} \varepsilon_\mathbf{k}}{N_\Xi-1}
\end{equation}
and is determined simply by choosing the closest value to the original dispersion
\begin{equation}
 \tilde\varepsilon_\mathbf{k} \equiv \mathrm{closest}(\Xi,\varepsilon_\mathbf{k})
\end{equation}
With a sufficiently dense grid $\Xi$, the approximation becomes negligible. We present the approximate $\tilde\varepsilon_\mathbf{k}$ we used in our calculations in comparison to the exact dispersion in Fig.~\ref{fig:epsk}.

\begin{widetext}
\subsection{Multiple precision algebra and regulators} \label{sec:gmp}

To illustrate the need for multiple precision algebra, we focus here on the simplest example, which is the second order diagram. The Matsubara summations here can be easily carried out by hand
\begin{eqnarray}
 && {\cal D}_\mathbf{k}(i\omega) =\\ \nonumber 
 && \;\; (-1) \sum_{i\omega',i\nu}\sum_{\mathbf{k}',\mathbf{q}}
   G^\mathrm{HF}_{0,\mathbf{k}'}(i\omega')
   G^\mathrm{HF}_{0,\mathbf{k}'+\mathbf{q}}(i\omega'+i\nu)
   G^\mathrm{HF}_{0,\mathbf{k}-\mathbf{q}}(i\omega-i\nu) \\ \nonumber
 && = \;\; (-1) \sum_{i\omega',i\nu}\sum_{\mathbf{k}',\mathbf{q}} 
   \frac{1}{i\omega'-\varepsilon_\mathbf{k'}}
   \frac{1}{i\omega'+i\nu-\varepsilon_{\mathbf{k'}+\mathbf{q}}}
   \frac{1}{i\omega-i\nu-\varepsilon_{\mathbf{k}-\mathbf{q}}}  \\ \nonumber
 && = \;\;\sum_{\mathbf{k}',\mathbf{q}}    
 \frac{
     n_F(\varepsilon_\mathbf{k'})n_B(\varepsilon_{\mathbf{k'}+\mathbf{q}}-\varepsilon_\mathbf{k'})   
 +   n_F(\varepsilon_\mathbf{k'})n_F(-\varepsilon_{\mathbf{k}-\mathbf{q}})
 -   n_F(\varepsilon_{\mathbf{k'}+\mathbf{q}})n_B(\varepsilon_{\mathbf{k'}+\mathbf{q}}-\varepsilon_\mathbf{k'})   
 -  n_F(\varepsilon_{\mathbf{k'}+\mathbf{q}})n_F(-\varepsilon_{\mathbf{k}-\mathbf{q}})
 }{i\omega-\varepsilon_{\mathbf{k'}+\mathbf{q}}-\varepsilon_{\mathbf{k}-\mathbf{q}}+\varepsilon_\mathbf{k'}}   
\end{eqnarray}
We see that the final result has four terms in total, and that the two terms featuring $n_B$ diverge as $\mathbf{q}\rightarrow 0$, or equivalently as $t\rightarrow 0$, i.e. $\varepsilon_\mathbf{k}\rightarrow0,\forall \mathbf{k}$. Nevertheless, the contribution of the diagram is \emph{finite} as the following limit is well-defined
\begin{equation}
\lim_{\varepsilon\rightarrow0}
\left( n_F(0)n_B(\varepsilon)  -n_F(\varepsilon)n_B(\varepsilon)    \right) 
   = \frac{1}{4}      
\end{equation}
However, in numerical implementation one cannot simply let $\varepsilon\rightarrow 0$ in the above expression as $n_B$ becomes ill-defined. We find the solution in adding small shifts in the symbolic expression. At second order, it suffices to associate a small shift $\epsilon$ to $\varepsilon_{\mathbf{k}'}$.
\begin{eqnarray}
 && {\cal D}_\mathbf{k}(i\omega) \approx
 \sum_{\mathbf{k}',\mathbf{q}}    
   \frac{1
   }{i\omega-\varepsilon_{\mathbf{k'}+\mathbf{q}}-\varepsilon_{\mathbf{k}-\mathbf{q}}+\varepsilon_\mathbf{k'}+\epsilon}   
   \times \\ \nonumber
&& \times\Bigg[  
      n_F(\varepsilon_\mathbf{k'}+\epsilon)n_B(\varepsilon_{\mathbf{k'}+\mathbf{q}}-\varepsilon_\mathbf{k'}-\epsilon)   
   +  n_F(\varepsilon_\mathbf{k'}+\epsilon)n_F(-\varepsilon_{\mathbf{k}-\mathbf{q}})
   -  n_F(\varepsilon_{\mathbf{k'}+\mathbf{q}})n_B(\varepsilon_{\mathbf{k'}+\mathbf{q}}-\varepsilon_\mathbf{k'}-\epsilon)   
   -  n_F(\varepsilon_{\mathbf{k'}+\mathbf{q}})n_F(-\varepsilon_{\mathbf{k}-\mathbf{q}})
   \Bigg]
\end{eqnarray}
That solves the problem as $n_B$ will no longer be ill-defined even when $\mathbf{q}=0$. However, depending on the size of $\epsilon$ and $\beta$, the two problematic terms may become large. Consider $\epsilon=10^{-20}$ and $\beta=1$. In that case the terms featuring $n_B$ can become as big as $10^{20}$. The subtraction of two numbers of size $10^{20}$ that are different only by $\frac{1}{4}$ will fail if performed in standard (double) precision, as it handles only up to $\sim 16$ digits. While in the case of second order diagram one can clearly use a larger $\epsilon$ and avoid any problems, at higher perturbation orders there will be products of several diverging $n_B$, multiplied also with expressions of the type $1/0$, and ever larger shifts would be needed; increasing the numerical shifts would eventually start introducing noticeable systematic error. The solution is then to use larger floating point data types that can store more digits. In our implementation we use GNU Multiple Precision Arithmetic (GMP) C++ library and its python wrapper GMPY2 and use floating point type of 350 bits, and we keep the shifts perturbation order dependent, $\sim 10^{-12+N}$.

\end{widetext}

\subsection{Monte Carlo application to local self-energy} \label{sec:local_mc}

We also devise an algorithm to treat directly the local self-energy. This algorithm relies on rewriting the diagrams in real space. 
In notation analogous to Eq.~\ref{eq:diag_contrib}, the contribution of a general real space diagram has the following form
\begin{eqnarray} \label{eq:realspacediag}
 && {\cal D}_{\mathbf{i}_0\mathbf{i}_{N}}(i\omega) = (-1)^{N_b}\sum_{\mathbf{i}_1...\mathbf{i}_{N-1}}\sum_{i\Omega_1..i\Omega_M} \times \\ \nonumber
 && \;\;\;\;\;\;\;\;\;\;\times \int d\varepsilon_1...d\varepsilon_{2N-1} \prod_\gamma \frac{\rho_{\mathbf{r}(\gamma;\mathbf{i}_0...\mathbf{i}_{N})}(\varepsilon_\gamma)}{
  \sum_{(s,j)\in {\cal K}_\gamma} s i\Omega_j -\varepsilon_\gamma}
\end{eqnarray}
where $\mathbf{i}_i$ denote the lattice-sites where the interaction vertices are positioned (the first and last are the external site indices). The energy integrals come from the Hilbert transform
\begin{equation}
 G_\mathbf{r}(i\omega) = -\frac{1}{\pi} \int d\epsilon \frac{\mathrm{Im}G_\mathbf{r}(\varepsilon+i0^+)}{i\omega-\varepsilon}
\end{equation}
and 
\begin{eqnarray} \nonumber
 \rho_\mathbf{r}(\varepsilon) &=& -\frac{1}{\pi}\mathrm{Im}G_\mathbf{r}(\varepsilon+i0^+) \\ \nonumber
                              &=& -\frac{1}{\pi}\mathrm{Im}\sum_\mathbf{k} e^{i\mathbf{k}\cdot\mathbf{r}}G_\mathbf{k}(\varepsilon+i0^+) \\ 
                              &=& \sum_\mathbf{k} e^{i\mathbf{k}\cdot\mathbf{r}} \delta_{\varepsilon,\varepsilon_\mathbf{k}} \\ \nonumber
                              &=& 2\sum_{ 0<k_x,k_y<\pi } \Big(\cos(\mathbf{k}\cdot\mathbf{r})+\cos(\mathbf{k}\sigma^z\mathbf{r})\Big) \delta_{\varepsilon,\varepsilon_\mathbf{k}} 
\end{eqnarray}
where $\mathbf{k}\sigma^z\mathbf{r}=k_xr_x-k_yr_y$. The above can be evaluated numerically to high precision. It is important to note that
\begin{eqnarray}
 \int d\varepsilon \rho_{\mathbf{r}=(0,0)}(\varepsilon) = 1 \\
 \int d\varepsilon \rho_{\mathbf{r}\neq(0,0)}(\varepsilon) = 0
\end{eqnarray}
Now note that only $\rho$ actually depends on the choice of lattice sites. We rewrite the expression in a way that is more revealing
\begin{eqnarray} \nonumber
 && {\cal D}_{\mathbf{i}_0\mathbf{i}_{N}}(i\omega) = (-1)^{N_b}\sum_{i\Omega_1..i\Omega_M}\int d\varepsilon_1...d\varepsilon_{2N-1} \prod_\gamma \times \\ 
 && \;\;\;\times 
  \frac{1}{\sum_{(s,j)\in {\cal K}_\gamma} s i\Omega_j -\varepsilon_\gamma} \sum_{\mathbf{i}_1...\mathbf{i}_{N-1}}\rho_{\mathbf{r}(\gamma;\mathbf{i}_0...\mathbf{i}_{N})}(\varepsilon_\gamma)
\end{eqnarray}
For a given choice of $\varepsilon$'s and $\mathbf{i}$'s, this is formally the same as what we had in Eq.~\ref{eq:diag_contrib} in the main text. A completely analogous symbolic algebra algorithm can be used to resolve the Matsubara summations, but the results will be different. The difference from the $\mathbf{k}$-space case is that all the $\varepsilon$'s are now independent, which will lead to different analytical expressions for each diagram. The final expressions will, however, have the same general form (Eq.~\ref{eq:sum_of_poles} and Eq.~\ref{eq:amp_final} in the main text), yet slightly simplified: now one obtains only simple poles because no two Green's functions are identical, i.e. $m_\gamma=1,\forall\gamma$. In fact, even in the $\mathbf{k}$-space case, higher order poles appear only in dressed diagrams - a skeleton series would not have this feature. After the analytical summation of the Matsubara frequencies, the remaining expression to be evaluated has the form
\begin{eqnarray} \nonumber
 && {\cal D}_{\mathbf{i}_0\mathbf{i}_{N}}(z) = (-1)^{N_b}\int d\varepsilon_1...d\varepsilon_{2N-1} \sum_\kappa \frac{A_\kappa}{z-\omega_\kappa}\times \\
 && \;\;\;\;\;\times 
  \prod_\gamma \sum_{\mathbf{i}_1...\mathbf{i}_{N-1}}\rho_{\mathbf{r}(\gamma;\mathbf{i}_0...\mathbf{i}_{N})}(\varepsilon_\gamma)
\end{eqnarray}
where $A$ and $\omega$ implicitly depend on $\varepsilon_1...\varepsilon_{2N-1}$.
The remaining variables to be summed over now include both the energies $\varepsilon$ and the lattice sites $\mathbf{i}$. Note however, that  $A$ and $\omega$ do \emph{not} depend on the $\mathbf{i}$'s, so recalculating them for each configuration of $\mathbf{i}$'s would be inefficient. We are immediately inclined to use $\prod_\gamma \sum_{\mathbf{i}_1...\mathbf{i}_{N-1}}\rho_{\mathbf{r}(\gamma;\mathbf{i}_0...\mathbf{i}_{N})}(\varepsilon_\gamma)$ as the weight for Monte Carlo over the space of $\varepsilon$'s. We recall the general expression
\begin{equation}
   \frac{\int f(x) w(x) dx}{\int w(x) dx} = \frac{\sum_{x\in \mathrm{MC}(|w|)} f(x) \mathrm{sgn}(w(x)) }{ \sum_{x\in \mathrm{MC}(|w|)} \mathrm{sgn}(w(x)) }
\end{equation}
 where $\mathrm{MC}(|w|)$ is Markov chain constructed with respect to $|w|$ as the weight.
Therefore it is necessary that the overall integral of our weight function is known and non-zero. However, this will only be the case if $\mathbf{i}_0=\mathbf{i}_N$. First, the integrals over our proposed weight decouple
\begin{eqnarray} \nonumber
 &&\int d\varepsilon_1...d\varepsilon_{2N-1} \prod_\gamma \sum_{\mathbf{i}_1...\mathbf{i}_{N-1}}\rho_{\mathbf{r}(\gamma;\mathbf{i}_0...\mathbf{i}_{N})}(\varepsilon_\gamma) \\
 &&\;\;\;\;\;\; = \sum_{\mathbf{i}_1...\mathbf{i}_{N-1}} \prod_\gamma \int d\varepsilon_\gamma \rho_{\mathbf{r}(\gamma;\mathbf{i}_0...\mathbf{i}_{N})}(\varepsilon_\gamma)
\end{eqnarray}
We see that the only contribution comes from the choice $\mathbf{i}_0=\mathbf{i}_1=...=\mathbf{i}_N$ in which case $\mathbf{r}(\gamma;\mathbf{i}_0...\mathbf{i}_{N})=(0,0), \forall\gamma$, and so each integral over energy equals 1, and the total integral of the weight is also equal 1. Otherwise, if $\mathbf{i}_0\neq\mathbf{i}_N$, there will always be at least one non-local $\rho_\mathbf{r}(\varepsilon)$ involved, the integral of which is 0. Therefore, the proposed weight has total integral zero for any non-local self-energy component and cannot be used in this purpose. Nevertheless, one can use it for calculating the local self-energy. Furthermore, in a local problem, e.g. Anderson impurity\cite{hewson1993}, this scheme can be used straight-forwardly without the summations over lattice sites. We use it in our Anderson impurity benchmark below.

%

\subsection{Diagram topologies} \label{sec:diagram_topologies}

\begin{figure}[ht!]
 \includegraphics[width=3.1in, trim=0 8cm 0 7cm, clip]{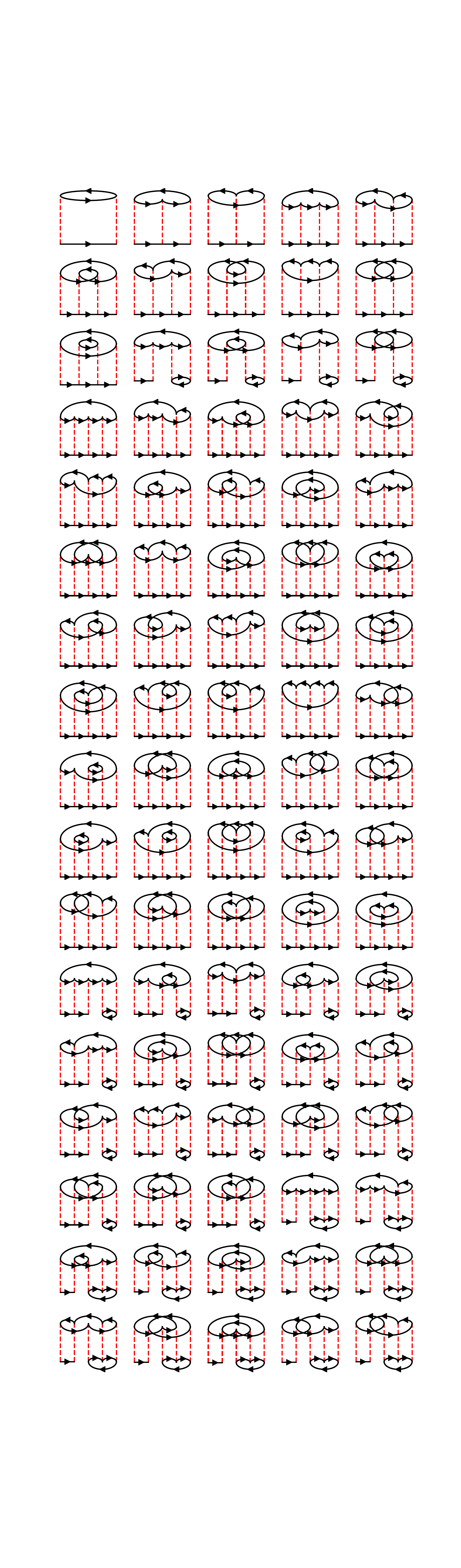}
 \caption{Hartree-shifted self-energy series up to 5th order.  The numbers of diagrams per order are 1,2,12,70,515,... starting from the second order, respectively. 
 }
 \label{fig:diags}
\end{figure}

In Fig.\ref{fig:diags} we present all the topologies of the interaction-expansion diagrams up to order 5. Full lines are the Hartree-shifted bare propagators, and the dashed lines are interactions. All the drawn diagrams went into calculation of the self-energy in the Fig.~\ref{fig:32x32w}.

\section{Benchmark} \label{sec:benchmark}
Here we benchmark our method in the following cases:
\begin{itemize}
 \item atomic limit against analytic result
 \item 4-site Hubbard chain against exact diagonalization (ED) \cite{strandPyED}
 \item 4x4 lattice against numerically exact Rubtsov algorithm, continuous-time interaction-expansion quantum Monte Carlo (CTINT)\cite{rubtsov2004,rubtsov_prb_2005,ParcolletCPC2014}
 \item single Anderson impurity problem against the approximative NRG\cite{ZitkoPRB2009}
 \item 32x32 lattice against imaginary-time diagrammatic Monte Carlo, $\Sigma$Det, up to 8th order in perturbation theory\cite{moutenet2018,fedor_2017}
\end{itemize}

\subsection{Atomic limit}

We start by benchmarking our method in the case of the half-filled Hubbard atom. It corresponds to setting $t=0$, $\mu=U/2$ (and $n_\sigma=0.5$ in the definition of the Hartree-shifted bare propagator). As there is no longer $\mathbf{k}$-dependence in the dispersion, the $\mathbf{k}$-sums now reduce to a single term, and each diagram needs to be evaluated only once, for $\varepsilon_\mathbf{k}=0$. As explained in Appendix~\ref{sec:gmp}, this cannot be done straight-forwardly because it would lead to divergent terms in the analytical expression, namely of the form $n_B(0)$ and $1/0$ (see Eq.~\ref{eq:amp_final} and the example Eq.\ref{eq:symres_example}). The numerical treatment boils down to adding small shifts to a certain number of $\varepsilon$'s at the symbolic level so that zeros are avoided in the arguments of $n_B$ and denominators of fractions, and only then letting the original $\varepsilon$'s go to zero (say, $\varepsilon_{\mathbf{k}_2-\mathbf{q}_1} \rightarrow \varepsilon_{\mathbf{k}_2-\mathbf{q}_1} + \zeta$, $\;\;\varepsilon_{\mathbf{k}_1} \rightarrow \varepsilon_{\mathbf{k}_1} + 2\zeta$, and so on, simultaneously across all terms in a given diagram; the shifts are integer multiples of $\zeta$ which we set depending on perturbation order $\zeta = 10^{-12+N}$; the choice of $\varepsilon$'s to be shifted is non-unique). This will a priori lead to systematic numerical error and here we check whether the numerical treatment is satisfactory (the atomic limit is the worst case scenario in this respect).

First, we recall the analytical expression for the self-energy beyond the Hartree shift
\begin{equation}
 \Sigma^{(\mathrm{HF})}(i\omega_n) = \frac{U^2}{4}\frac{1}{i\omega_n}
\end{equation}
It can be shown that this expression corresponds to the second order diagram in the $U$-series written down in terms of the Hartree-shifted bare propagator. The contribution of higher orders is zero ``order by order'', but individual higher-order diagrams are not necessarily zero. Therefore, it is a stringent check of our method to show that the higher orders truly cancel. 

\begin{figure}
 \includegraphics[width=3.2in]{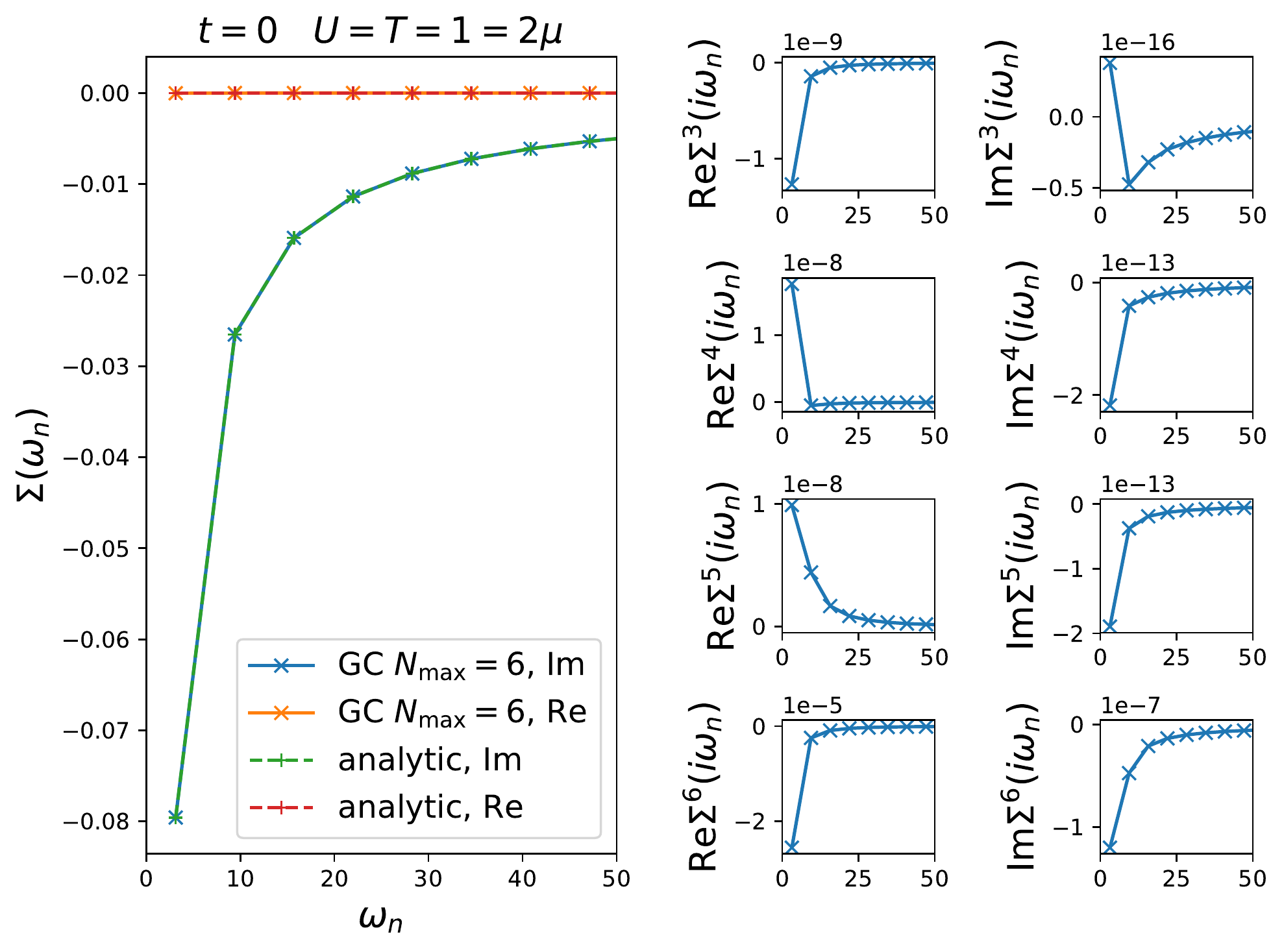}
 \caption{ Benchmark in the case of an isolated Hubbard atom at half-filling. Big panel: Our method (GC) is compared to the analytical expression. Smaller panels on the right: self-energy contributions order-by-order; the only contribution comes from the second order diagram.
 }
 \label{fig:atomic}
\end{figure}

We present the results in Fig.~\ref{fig:atomic}. We evaluate all the diagrams up to and including the 6th order, at a fixed $U=T=1$. The total series is in excellent agreement with the analytical result (big panel). On the smaller panels on the right, we examine the contributions order by order ($\Sigma^N$ denotes contribution at order $N$). Indeed, the only contribution comes from the second order diagram, while the contributions of higher orders are negligible. However, the numerical error grows with approaching the real axis, and with growing order. The real part of self-energy coming from the 6th order diagrams already reaches $10^{-5}$. This is expected, as we use bigger numerical shifts in higher-order diagrams. Alternatively, one would need to drastically increase the floating-point precision in the evaluation of higher order diagrams, which is not suitable for lattice computations, so we do not consider this approach; rather, we keep the floating-point precision fixed across orders.

In the atomic limit, the real frequency self-energy cannot be reliably extracted from our method. This is, however, a somewhat pathological case where the self-energy is a single simple pole at $\omega=0$. Due to numerical shifts and cutting the series at finite order, our numerical self-energy here is composed of multiple poles of various orders at various small frequencies $\sim \zeta$. Very close to the real axis, these numerical artifacts become apparent, and the method is of little use.

\begin{figure}
 \includegraphics[width=3.2in, page=1]{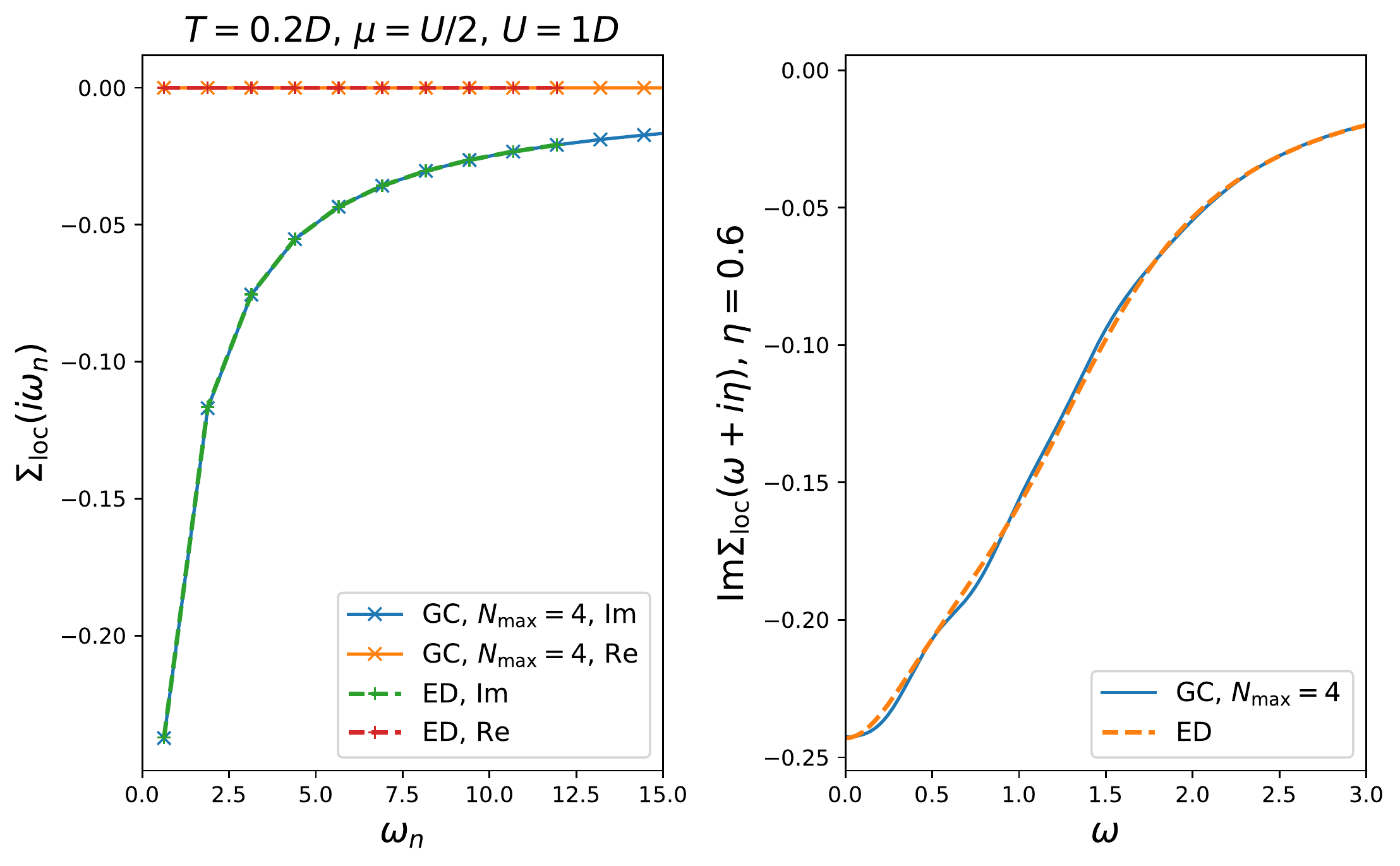}
 \includegraphics[width=3.2in, page=2]{Figure2x2}
 \caption{Benchmark in the case of the 4-site cyclic Hubbard chain at half-filling.
 }
 \label{fig:2x2}
\end{figure}

\subsection{4-site Hubbard chain}

Next, we benchmark our method in the case of the half-filled 4-site cyclic Hubbard chain at temperature $T=0.2D=0.8t$ (note that the actual half-bandwidth in this case is $2t$). This small system can be solved using exact diagonalization (ED). In our method, the $\mathbf{k}$-summations go over only 4 points and can be performed fully, so we denote our method GC (gray code). In this case we go up to order 4 (due to particle-hole symmetry, the order 5 does not contribute, but order 6 we cannot fully sum).

We present our result on Fig.~\ref{fig:2x2}. The agreement is excellent at $U=1D$, yet at $U=1.5D$ higher orders become important.

Similarly to the atomic limit, the self-energy in the 4-site chain is comprised from a relatively small number of poles on the real axis, and does not form a smooth frequency spectrum. On the other hand, having that $\varepsilon_\mathbf{k}$ takes on only three distinct values ($-0.5,0,0.5$), our method can yield poles only at frequencies which are integer multiples of $1/2$ (plus/minus small numerical shifts).
The immediate question is then: how does one recover the correct self-energy even with an infinite self-energy series? One would expect the poles in self-energy to appear at various different frequencies and even move continuously with increasing $U$, yet our analytical expression seemingly does not support that. 
The answer is that all the higher order poles ultimately merge into (shifted) simple poles through
\begin{equation}\label{eq:taylor_series}
\sum_{k=1}^{\infty} \frac{a^{k-1}}{z^k} = \frac{1}{z-a}
\end{equation}
and that way recover the correct physical result.
Note, however, that the principle part of the Laurent series Eq.\ref{eq:taylor_series} cut at a finite order, no longer resembles a simple pole at $\eta \lesssim a$, irrespective of the maximum order in the series.
Therefore, it makes no sense to look at $\Sigma(\omega+i\eta)$ results at small $\eta$. One reasoning is that we should take $\eta$ proportional to the distance between the poles we get, which is in this case $0.5$. We therefore compare our result to ED at $\eta=0.6$ which is just below the first Matsubara frequency $\pi T$ and find similarly good agreement as on the imaginary axis.

Again, our method cannot be used to reliably extract discrete spectra on the real axis. Fitting the result at $\eta=0.6$ to a causal and piecewise constant spectrum on the real axis does reproduce the correspondingly binned ED result, but the detailed pole structure cannot be inferred.

\begin{figure}[ht!]
 \includegraphics[width=3.2in, page=1]{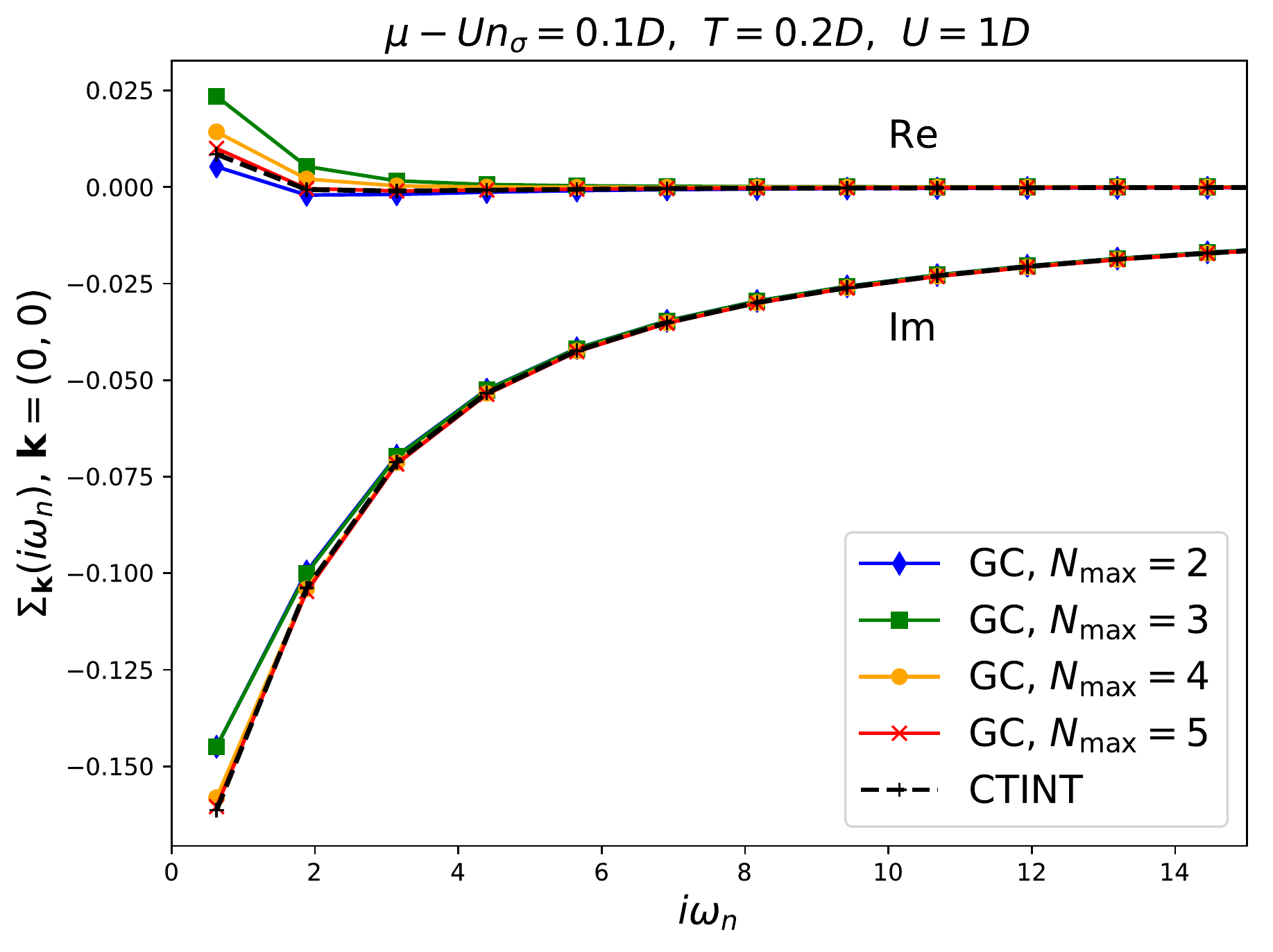}
 \includegraphics[width=3.2in, page=2]{Figure4x4}
 \caption{Benchmark of the method in the case of $4\times 4$ cyclic Hubbard cluster.
 }
 \label{fig:4x4}
\end{figure}

\begin{figure}[ht]
 \includegraphics[width=3.2in]{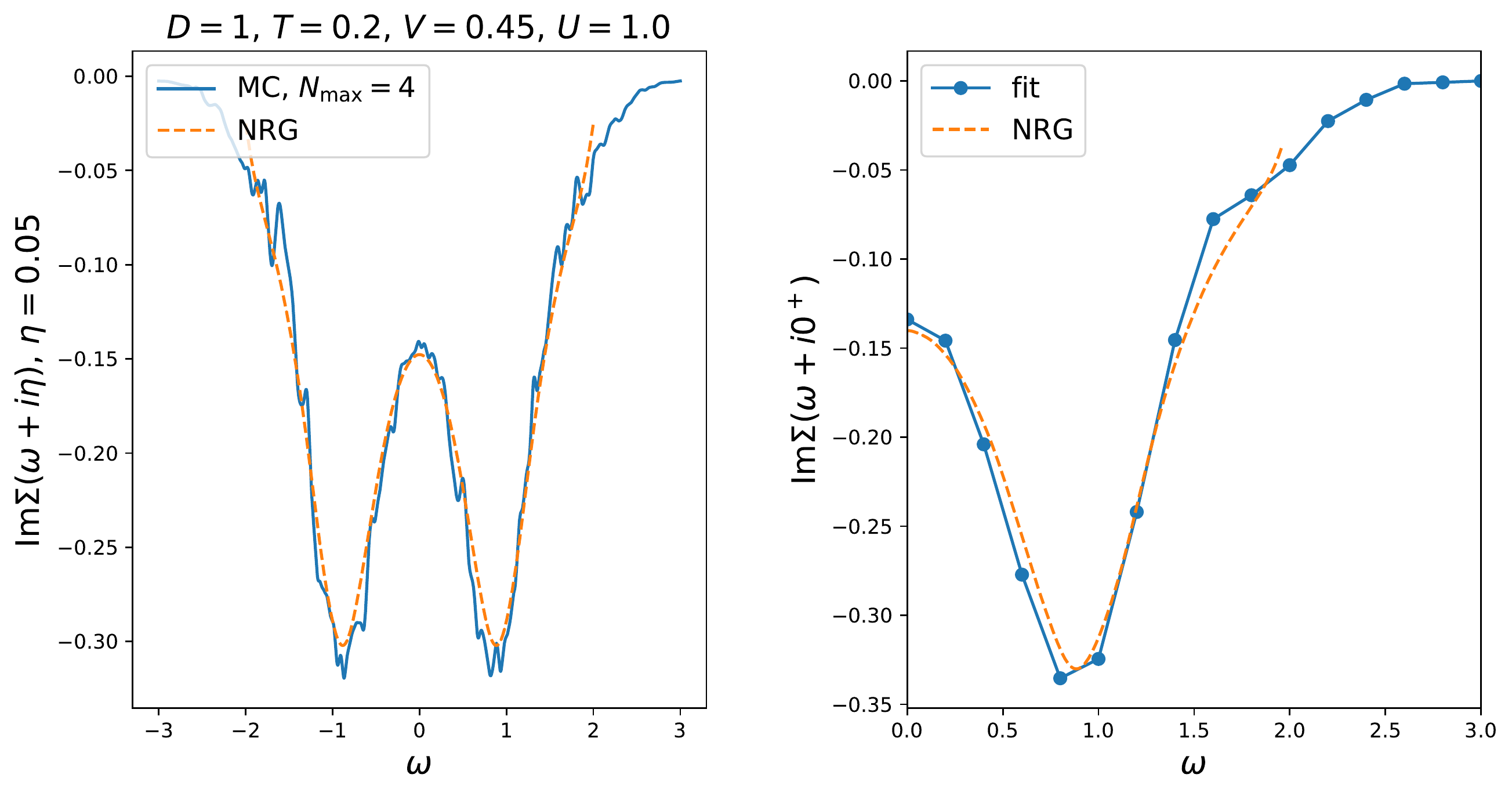}
 \caption{Benchmark of our method in the case of single-impurity Anderson model with a semi-circular bath.
 }
 \label{fig:aim}
\end{figure}

\begin{figure*}[ht!]
 \includegraphics[width=6.4in, trim=0cm 0 0cm 0, clip]{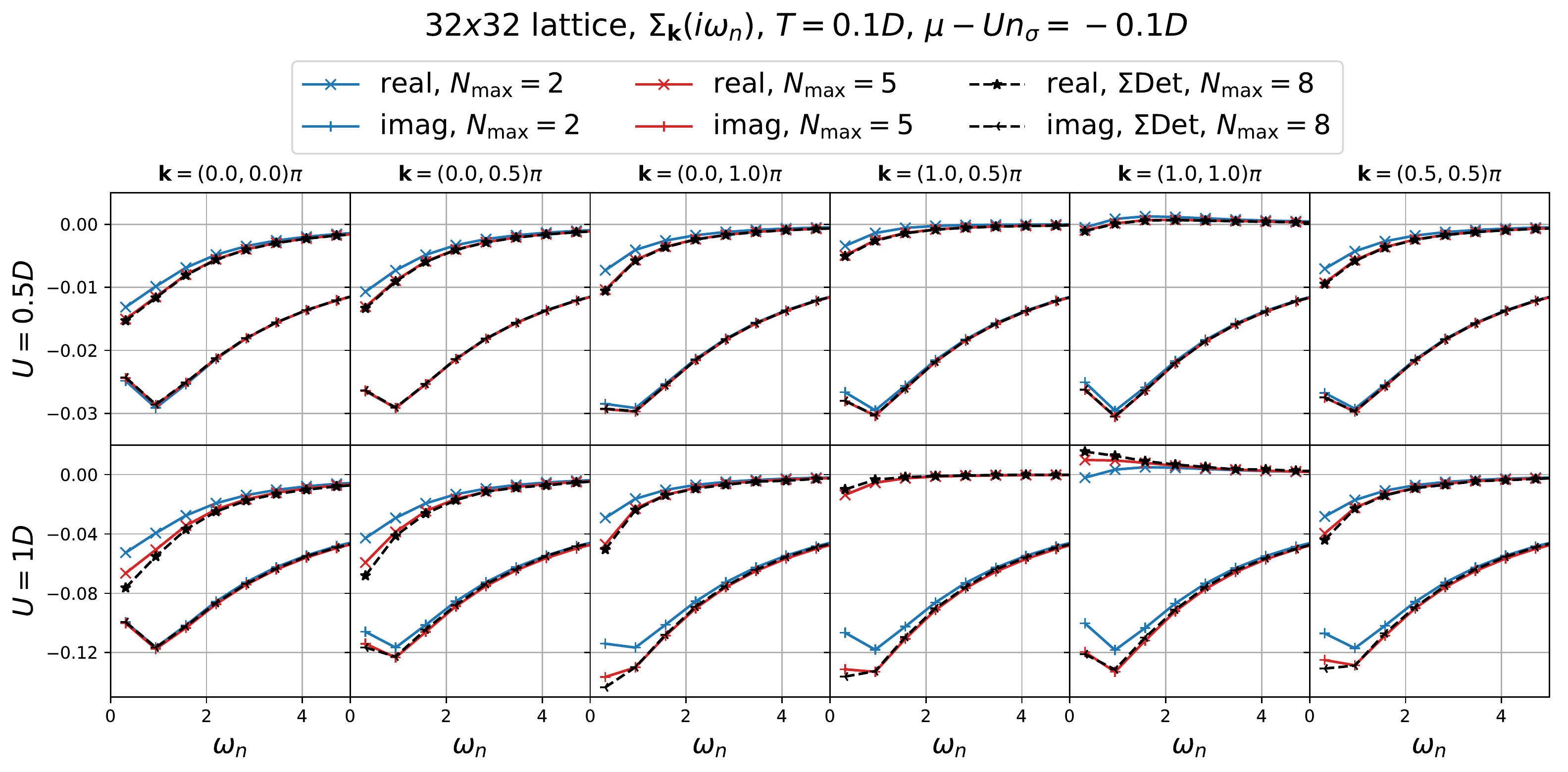}
 \caption{Matsubara self-energy on the 32x32 Hubbard lattice: benchmark against the $\Sigma$Det method at 8th order.}
 \label{fig:32x32iw}
\end{figure*}

\subsection{4x4 lattice}

We now turn to the $4\times 4$ cyclic Hubbard cluster. This system cannot easily be solved with ED, so we use the Rubtsov algorithm continuous-time interaction expansion Monte Carlo (CTINT) which is numerically exact. However the comparison can now only be made on the imaginary axis. In our method, full $\mathbf{k}$-summations can be performed up to order 5. 

In Fig.~\ref{fig:4x4} we show the results at $\mu-Un_\sigma=0.1D$, $T=0.2D$, $\mathbf{k}=(0,0)$. Additionally, we show the GC results for different perturbation order cutoffs $N_\mathrm{max}=2...5$. At $U=1D$ the agreement is excellent and the perturbation series seems converged at order 5. At $U=1.5D$ the agreement is solid, but 5th order still makes a sizeable contribution. 

As for the real-frequency spectrum, there is a similar problem as in the 4-site chain case - dispersion now assumes only the values $(\pm1,\pm0.5,0)$, and again one obtains poles only at integer multiples of $0.5$. The spectrum is expected to be discrete and dense, and any kind of fit to $\eta\sim0.5$ result is likely to miss details of it. Our method is suitable only for continuous spectra, as we will show in the following sections.

\subsection{Anderson impurity}

To test our method in the continuous spectrum case, we start with the simplest possible model: the Anderson impurity model with a semi-circular bath. We consider only the ph-symmetric case. The Hartree-shifted bare propagator is given by
\begin{equation}
 G^{\mathrm{HF}}_0(z) = \frac{1}{z-\Delta(z)}
\end{equation}
and the hybridization function
\begin{eqnarray}
 \Delta(z) &=& V^2\int  d\varepsilon \frac{\rho(\varepsilon)}{z-\varepsilon}  \\
 \rho(\varepsilon) &=& \theta( D - |\varepsilon|) 2 \sqrt{ D^2 - \varepsilon^2}/(\pi D^2)
\end{eqnarray}
where $V^2$ sets the norm, and $D$ sets the width of $\mathrm{Im}\Delta(\omega)$.

This model can be solved approximately using numerical renormalization group (NRG). NRG yields the self-energy directly on the real axis.

In our method, we utilize the real space algorithm introduced in Section \ref{sec:local_mc}, with the important simplification that there are no sums over lattice sites. We discretize the energy (200 points between -1 and 1), and perform Monte Carlo integration for the $\varepsilon$ integrals using the product $\prod_\gamma \rho(\varepsilon_\gamma)$ as the weight. 

A priori, now we should be able to approach the real axis to around $\eta\sim1/100$. However, the statistical error now also plays the role, and we find that $\mathrm{Im}\Sigma(\omega+i\eta)$ becomes noisy below $\eta\sim0.05D$. Nevertheless, this should be sufficient to resolve all the details of the spectrum. We compare our results to NRG at $\eta=0.05$ and find excellent agreement (Fig.~\ref{fig:aim}). Note that we do not impose the ph-symmetry, but the result is ph-symmetric apparently within the level of noise in the curve. 
Next, we fit our result at $\eta=0.05D$ to a ph-symmetric piecewise-trapezoid spectrum on the real-axis with resolution $\sim 0.1$ and compare to the NRG result on the real axis. The agreement is excellent, and the resolution is sufficient to capture all the features in $\mathrm{Im}\Sigma(\omega+i0^+)$.

\subsection{32x32 lattice} \label{sec:bench_32x32}

Finally, we benchmark our method in the 32x32 Hubbard lattice case. The best available result is that of the imaginary-time $\Sigma$Det diagrammatic Monte Carlo calculation, performed up to 8th order. We compare the two methods on the Matsubara axis in Fig.\ref{fig:32x32iw}.

At $U=0.5D$ the agreement is excellent, and the calculation is clearly converged by order 5, but clearly not by order 2.

At $U=1D$ higher orders still contribute, and there is a bit of discrepancy at low frequency. From the real-frequency results (Fig.~\ref{fig:32x32w} in the main text), however, it is clear that the self-energy is qualitatively converged, although some corrections are expected with inclusion of higher orders. 

We do not benchmark using $U=1.5$ data, as in that case the higher orders are expected to contribute more, and results are not expected to coincide.

\bibliography{refs_merged.bib}
\bibliographystyle{apsrev4-1}


\end{document}